\documentclass[aps,prb,twocolumn,groupedaddress,showpacs,preprintnumbers,amsmath]{revtex4-1}
\usepackage{epsfig,amsmath,amssymb,graphics,color,calc}
\usepackage{braket}
\usepackage{mathtools}

\newcommand{\de}{\textrm{d}}

\begin{document}

\title{Anomalous dynamics in the ergodic side of the Many-Body Localization transition and the glassy phase of Directed Polymers in Random Media}

\author{G. Biroli\textsuperscript{1,2}, and M. Tarzia\textsuperscript{3,4}}

\affiliation{\textsuperscript{1}\mbox{Laboratoire de Physique de l'Ecole Normale Sup\'erieure, ENS}\\
\textsuperscript{2}\mbox{Universit\'e PSL, CNRS, Sorbonne
Universit\'e, Universit\'e de Paris, F-75005 Paris, France}\\ 
\textsuperscript{3} \mbox{LPTMC, CNRS-UMR 7600, Sorbonne Universit\'e, 4 Pl. Jussieu, F-75005 Paris, France}
\textsuperscript{4} \mbox{Institut  Universitaire  de  France,  1  rue  Descartes,  75231  Paris  Cedex  05,  France}
}

\begin{abstract}
Using the non-interacting Anderson tight-binding model on the Bethe lattice as a toy model for the many-body quantum dynamics, we propose a novel and transparent theoretical explanation of the anomalously slow dynamics that emerges in the bad metal 
phase preceding the Many-Body Localization transition.
By mapping the time-decorrelation of many-body wave-functions onto Directed Polymers in Random Media, we show the existence of a glass transition within the extended regime separating a metallic-like phase at small disorder, where delocalization occurs on an exponential number of paths, from a bad metal-like phase at intermediate disorder, where resonances are formed on rare, specific, disorder dependent site orbitals on very distant generations. 
The physical interpretation of subdiffusion and non-exponential relaxation emerging from this picture is complementary to the Griffiths one, although both scenarios rely on  the presence of heavy-tailed distribution of the escape times.
We relate the dynamical evolution in the glassy phase to the depinning transition of Directed Polymers, which results in macroscopic and abrupt jumps of the preferred delocalizing paths when a parameter like the energy is varied, and produce a singular behavior of the overlap correlation function between eigenstates at different energies. By comparing the quantum dynamics on loop-less Cayley trees and Random Regular Graphs we discuss the effect of loops, showing that in the latter slow dynamics
and apparent power-laws extend on a very large time-window but are eventually cut-off on a time-scale that diverges at the MBL transition.
\end{abstract}

\pacs{}

\maketitle

\section{Introduction} 

The field of Many-Body Localization (MBL) started about $10$ years ago with the work of Ref.~[\onlinecite{BAA}]
which studied the stability of the Anderson insulator with respect to the addition of interactions via 
the so-called self-consistent Born approximation for the one-particle Green's functions, showing
that isolated disordered many-body systems can fail to thermalize even at finite energy density if the 
disorder is strong enough.
MBL is a purely quantum phenomenon which occurs due to Anderson localization in the
Fock space as the result of the interplay of disorder, quantum fluctuations, and interactions,
and gives rise to a completely new mechanism for ergodicity breaking:\cite{reviewMBL,reviewMBL2,reviewMBL3,reviewMBL4,reviewMBL5} 
Differently from (quantum or classical) integrable systems, the MBL
phase is stable to perturbations; Differently from (classical or quantum) phase transitions, 
it is not associated to any spontaneous symmetry breaking, and occurs  
without any signature in the static observables (and in isolated systems only);
The MBL state is also different from---although it shares some similarities with---classical or quantum
glasses; for instance, it establishes also in $1d$ models characterized by a not ``too'' complex energy landscape, 
and at infinite temperature.

In fact the existence of the MBL transition was predicted by Althsuler {\it et al.} in a seminal paper\cite{dot} 
already 10 years before the breakthrough of Refs.~[\onlinecite{BAA},\onlinecite{Gornyi}], by putting forward a paradigmatic representation of MBL in
terms of single-particle Anderson localization in Fock space.
In order to explain this analogy, 
let us focus on the following 
disordered Ising spin chain as a reference model 
\begin{equation} \label{eq:HMB}
	{\cal H}_{\rm MB} = \sum_{i=1}^N \left ( J_i \hat{\sigma}_i^z \hat{\sigma}_{i+1}^z + h_i \hat{\sigma}_i^z \right) + \sum_{i=1}^N \Gamma_i \hat{\sigma}_i^x \, ,
\end{equation}
with $h_i$ i.i.d in $[-h,h]$, for which the existence of the MBL transition has been proven rigorously\cite{LIOMS} (under the minimal assumption of absence of level attraction).
If one chooses as a basis the tensor product of the simultaneous eigenstates of the operators $\sigma_i^z$, 
the Fock space of the many-body Hamiltonian is a $N$-dimensional hyper-cube of ${\cal V} = 2^N$ sites.
The first part of ${\cal H}_{\rm MB}$ is by definition diagonal on this basis. Its diagonal elements correspond to correlated
random energies associated to the sites of the hyper-cube:
\begin{equation} \label{eq:random_energies}
	\langle \{ \sigma_i^z \} \vert \sum_{i=1}^N \left( J_i \hat{\sigma}_i^z \hat{\sigma}_{i+1}^z  + h_i \hat{\sigma}_i^z \right ) 
	\vert \{ \sigma_i^z \} \rangle
	= \epsilon (\{ \sigma_i^z \}) \, ,
\end{equation}
while the interacting part of the Hamiltonian induces single spin flips on the configurations $\{ \sigma_i^z \}$:
\begin{equation}
	\Gamma_i \hat{\sigma}_i^x \vert \sigma_1^z, \ldots, \sigma_i^z, \ldots, \sigma_N^z \rangle 
	= \Gamma_i \vert \sigma_1^z, \ldots, -\sigma_i^z, \ldots, \sigma_N^z \rangle \, ,
\end{equation}
and leads to hopping connecting neighboring sites of the hyper-cube.
The many-body quantum dynamics can then be thought as a tight binding model on a
very high-dimensional disordered lattice. In large spatial dimensions the neighbors of a given site are organized in
a peculiar way: their number grows very rapidly with the distance and short loops among them are rare. Since these 
are distinctive features of tree-like structures, the authors of Ref.~[\onlinecite{dot}] argued that the (non-interacting) Anderson model on the Bethe lattice, 
originally introduced and studied in Ref.~[\onlinecite{ATA}], can be used as a toy model for MBL (see also Refs.~[\onlinecite{Gornyi,BetheProxy1,BetheProxy2}] for a similar analysis and Ref.~[\onlinecite{scardicchioMB}] for a quantitative investigation of such mapping).

Based on this idea, the existence of three distinct regimes was suggested.
At strong disorder the many-body eigenfunctions are exponentially localized around some specific site orbitals
in the configuration space and are weak deformations of the non-interacting states: The system 
is a perfect insulator (i.e., conductivity is strictly zero) and is not ergodic, 
on-site energies are ``good'' quantum number 
(akin to the so-called local conserved quantities\cite{LIOMS,LIOMS1,LIOMS2,LIOMS3}), and the level statistics should be of Poisson type.
At weak disorder, instead, the wave-functions are extended over the whole accessible volume: 
the non-interacting states $\{ \sigma_i^z \}$ become effectively coupled to infinitely many other states
(i.e., to a continuum of energy levels), the system provides its own bath and behaves as a normal metal
(the statistics of energy levels should then be described by the GOE ensemble). Between the MBL and the 
metallic phase, the authors also predicted the possibility of the existence of an intermediate regime (nowdays called
the ``bad metal'') where the wave-functions might be delocalized but not ergodic: site orbitals in the
Fock space may only hybridize with an infinitesimal fraction of the accessible volume. 
This regime 
is expected to be characterized by highly heterogeneous transport and strong fluctuations (and, possibly, by anomalous level statistics).

A huge amount of work has been done on this subject in the latest years (see, e.g., Refs.~[\onlinecite{reviewMBL,reviewMBL2,reviewMBL3,reviewMBL4,reviewMBL5}] for recent reviews) and, as mentioned above, the existence of the MBL transition
has been even established at a mathematical level for some specific models under some minimal assumptions.\cite{LIOMS}
However, most of the studies have focused either on the MBL phase itself or on the 
transition point.

The interest on the delocalized side of the transition started recently, when it was observed
that the delocalized phase is actually very unusual:\cite{BarLev,reviewdeloc1} 
In fact it was found that in a broad range of parameter before MBL, 
transport is sub-diffusive and out-of-equilibrium relaxation toward thermal equilibrium
is anomalously slow and described by power-laws with exponents that gradually approach zero at the
transition.
These features appear as remarkably robust: They were 
observed in Ref.~[\onlinecite{daveBAA}], by solving numerically the equations obtained within the self-consistent Born approximation,\cite{BAA,Gornyi} 
in numerical simulations (using exact diagonalizations or time-dependent matrix product states) of disordered spin chains of moderate sizes,\cite{dave1,reviewdeloc1,BarLev,demler,alet,torres,luitz_barlev,doggen,evers} as well as 
 in recent experiments with cold atoms.\cite{experiments1,experiments2,experiments3}

An appealing phenomenological interpretation of these phenomena has been 
proposed in terms of Griffiths effects.\cite{griffiths,griffiths2,reviewdeloc1,BarLev}
The idea is that a system close to MBL is highly inhomogeneous (in real space) and is characterized by rare inclusions of the
insulating phase with an anomalously large escape time (i.e., anomalously small localization length). 
In $1d$ such insulating segments affect dramatically the dynamics, since quantum excitations have to go through broadly distributed 
effective barriers which act as kinetic bottlenecks and give rise to sub-diffusion 
and slow relaxation, in a way which is very similar to the trap model for glassy dynamics.\cite{trap}

However the Griffits picture is not completely satisfactory. In fact unusual transport and power-law relaxations
have been recently observed also in quasiperiodic $1d$ 
and disordered $2d$ systems, both in experiments\cite{experiments2,experiments3} and numerical
simulations,\cite{daverecent,mace,evers1,dave2d,doggen2d} while on general grounds one expects that Griffits effects should only give
a subdominant contribution when the potential is correlated and/or the dimension is larger than one.\cite{reviewdeloc1,griffiths2} It is therefore natural
to seek for other mechanisms that might hold beyond the specific case of $1d$ disordered systems.

In a recent paper,\cite{PLMBL} using the Anderson model on the Bethe lattice as a
pictorial representation for the many-body quantum dynamics (following Refs.~[\onlinecite{dot}] and~[\onlinecite{Gornyi,BetheProxy1,BetheProxy2,scardicchioMB}]), we proposed a possible complementary explanation of the slow and power-law-like 
relaxation observed in the bad metal phase directly based on quantum dynamics in the Fock space. 
More precisely our toy model is the tight-binding Hamiltonian for non-interacting spinless fermions (introduced in Ref.~[\onlinecite{ATA}]),
\begin{equation} \label{eq:H}
	{\cal H} = \sum_{x=1}^{\cal V} \epsilon_x \vert x \rangle \langle x \vert + t \sum_{\langle x,y \rangle} \big ( 
	\vert x \rangle \langle y \vert + \vert y \rangle \langle x \vert \big ) \, ,
\end{equation}
where the on-site random energies are taken as i.i.d. random variable uniformly distributed in $[-W/2,W/2]$, and
$\langle x,y \rangle$ denotes nearest-neighboring site on the Bethe lattice. In Ref.~[\onlinecite{PLMBL}] the Bethe lattice was taken as
a Random-Regular Graph (RRG), i.e., a random lattice which has locally a tree-like structure but has loops whose typical length
scales as $\ln {\cal V} \propto N$ and no boundary.\cite{RRG}
In the analogy with MBL discussed for the Hamiltonian (\ref{eq:HMB}), each site of the lattice should be interpreted as a many-body configurations, and on-site energies
as (extensive) random energies of the $N$-body interacting system, see Eq.~(\ref{eq:random_energies}).
Of course this analogy represents a drastic simplification of real systems, as one neglects the correlation between random energies as well as
the specific structure of the hyper-cube.
Moreover, we considered Bethe lattices of fixed connectivity (we set the total connectivity $k+1$ equal to $3$ throughout, as in Ref.~[\onlinecite{PLMBL}]), 
while the connectivity of the configuration space of the many-body system increases as $N \propto \ln {\cal V}$.
The counterpart of the MBL transition corresponds to Anderson localization which, for the non-interacting Hamiltonian~(\ref{eq:H}) we focus on, and for $k+1=3$ 
and $E=0$ (corresponding to the middle of the band, i.e., infinite temperature for the many-body system), 
takes place at $W_L \approx 18.2$, as obtained from 
previous studies of the transmission properties and dissipation propagation,\cite{ATA,garel,noi} and precisely determined in Ref.~[\onlinecite{tikhonov_critical}].

Defining suitable proxies of the correlation functions of local operators in real space of the original many-body problems (see Sec.~\ref{sec:dynamics}
for a detailed explanation), we
studied both the out-of-equilibrium and the (infinite temperature) equilibrium dynamics, 
and showed that in a broad region of the phase diagram the counterpart of the spin imbalance and of the equilibrium correlation function display slow relaxation and a power-law-like
behavior strikingly similar (at least qualitatively) to the one observed in the bad metal phase of many-body systems, with apparent dynamical exponents that evolve continuously 
with the disorder and approach zero at the localization transition (see also Ref.~[\onlinecite{scardicchio_dyn}] for a recent tightly related investigation).

Ref.~[\onlinecite{PLMBL}] is far from being the only work addressing MBL-related questions using Anderson in terms of localization on the Bethe lattice. In fact, inspired by the mapping of MBL onto single particle Anderson localization in a very high-dimensional 
space,\cite{dot,BetheProxy1,BetheProxy2,scardicchioMB} 
in the latest years an intense research activity has been devoted to establish the existence of a non-ergodic delocalized phase in the tight-binding
model on the Bethe lattice.\cite{noi,scardicchio1,ioffe1,ioffe3,mirlin,mirlinCT,levy,lemarie,Bethe,mirlintikhonov,scardicchio_sub,refael}  
As a matter of fact, the slow dynamics and power-law relaxation observed in Ref.~[\onlinecite{PLMBL}] emerge precisely in the
region $W_T \le W \le W_L$ where some of the previous studies have suggested that wave-functions might be extended but 
multifractal.\cite{noi,ioffe1,ioffe3}
Although recent results convincingly indicate that full ergodicity is eventually recovered on RRGs larger than a cross-over scale which diverges exponentially fast approaching the localization transition,\cite{Bethe,mirlin,lemarie,mirlintikhonov}
these observations suggest that the physical origin of the unusual slow and sub-diffusive dynamics observed in the bad metal phase
is tightly related to the apparent non-ergodic features of the spectral statistics that seem 
to emerge in the delocalized phase approaching the
localization transition. 
Here we come back to this problem. The main outcome of this paper is twofold. On the one hand,
using a mapping to directed polymer in random media, we obtain a clear, novel, and transparent physical interpretation of the unusual and slow dynamics
observed in many-body isolated disordered system approaching the MBL transition 
in terms of delocalization along rare, ramified, disorder-dependent paths in the Fock space.  
On the other hand, we show that the apparent non-ergodic features of the delocalized phase that 
have been found for Anderson localization on random regular graphs are the vestiges of the truly 
non-ergodic delocalized phase present on Cayley trees.\cite{mirlinCT,garel,garelDP} We thus offer an explanation as well as 
a quantitative theory of the cross-over phenomena (already extensively discussed in Ref.~[\onlinecite{Bethe}]) associated to the bad metal phase in RRGs.


{\it Summary of results}. In the next section, using the single-particle Anderson tight-binding model on the loop-less Cayley tree~(\ref{eq:H}) as a toy model, we map the problem of ergodicity for quantum dynamics to the one of Directed Polymers in Random Media (DPRM).\cite{DPRM,gabriel,garelDP} By analyzing the properties of the average free-energy of the DP, we show the existence of a sharp transition to a glassy phase where delocalization can only occur on few specific disorder-dependent paths. 
We discuss the manifestation of such freezing glass transition on the non-ergodicicity  
of the wave-functions, 
and in particular on the singular probability distribution of the Local Density of States (LDoS) at the root of the Cayley tree. In Sec.~\ref{sec:dynamics} we probe the dynamical evolution on the Cayley tree by measuring observables built as proxies for the imbalance and equilibrium correlation functions\cite{PLMBL} which display slow dynamics and power-laws in the glassy phase. 
We establish a quantitative connection between the emergence of the anomalously slow relaxation and the non-ergodic features of the spectral statistics. In particular we highlight the relationship between 
the depinning transition of the DP (i.e., of the preferred paths along which decorrelation can occur) in the glassy phase and the singular behavior of the overlap correlation function between eigenstates at different energies (which is essentially the Fourier transform to frequency space of the dynamical correlation function).
In Sec.~\ref{sec:loops}  we discuss the effect and the importance of loops by contrasting the dynamical evolution on the Cayley tree with the one observed on the RRG in Refs.~[\onlinecite{PLMBL}] and~[\onlinecite{scardicchio_dyn}], and offer an explanation as well as a quantitative theory of the cross-over phenomena associated to the non-ergodic-like behavior in RRGs.
Finally, in Sec.~\ref{sec:conclusions} we summarize the results found and discuss their physical implications, providing some perspectives for future work. Some technical aspects are discussed in details in the appendices~\ref{app:spectral}-\ref{app:return}.

\section{Delocalization in Fock space and Directed Polymers in Random Media} \label{sec:dprm}

\begin{figure}
\includegraphics[width=0.46\textwidth]{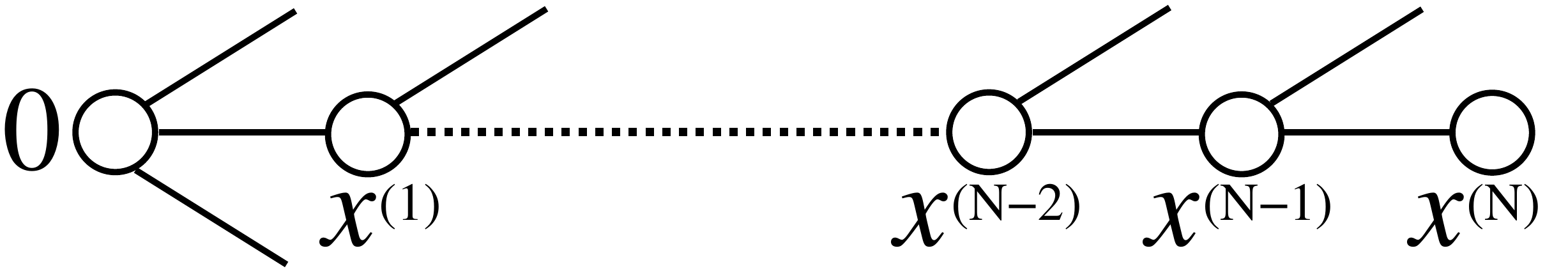}%
\caption{\label{fig:chain}
        Schematic representation of one of the $k^N$ paths on the Cayley tree ($k=2$ here) going from the root $0$ to a site of the boundary $x^{(N)}$ along which 
        the resolvent matrix element ${\cal G}_{x^{(N)},0}$ is computed according to Eq.~(\ref{eq:corr}).}
\end{figure}

Let us focus again on the many-body Hamiltonian~(\ref{eq:HMB}) as a reference model, and 
imagine to start at time $t=0$ from a random (infinite temperature) 
magnetization profile with $\sigma_i^z = \pm 1$ with probability $1/2$
(which corresponds to a specific site in 
the configuration space, 
$\vert \psi(t=0) \rangle = \vert x_0 \rangle$, whose random energy
is close to zero) 
and let us ask ourselves the following question: What is the probability 
that at large time the spin configuration has completely decorrelated from the initial one?
For a system of $N$ spins, this roughly corresponds to requiring that a finite fractions of them have flipped, i.e., that at time $t$ the system 
is found on a site $\vert x \rangle$ of the configuration space which is order $N$ steps (i.e., spin flips) away from the initial state.
Here we address this question by using the pictorial representation of the quantum many-body dynamics in terms of a single-particle Anderson model 
on tree like structures~\cite{dot,BetheProxy1,BetheProxy2,scardicchioMB,PLMBL}.
For sake of simplicity, we start by focusing on (loop-less) Cayley trees as the underlying lattice mimicking the Fock space, and address later the effect of loops by considering RRGs (see Sec.~\ref{sec:loops}). 

\subsection{From ergodicity of quantum dynamics to directed polymer in random media}
Let us consider a loop-less Cayley tree of $N$ generations of fixed connectivity $k+1=3$ and total number of sites 
${\cal V} \approx 2^N$,\cite{NCT} as our toy model for the Fock space.
A measure of the degree of ergodicity of quantum dynamics is the amount of spreading of a wave-function that is initially localized  at the root of the tree (labeled as $0$). More precisely, one wants to study the wave-function amplitude (at large times) on the sites of the boundary of the tree $\vert x^{{(N)}} \rangle$ (i.e., at distance of order $N \propto \ln {\cal V}$ away from the root) given the initial condition $\langle 0 \vert \psi (t) \rangle =1$ at $t=0$.
The time evolution of the wave-function at time $t$, 
$\vert \psi (t) \rangle$, 
can be written in terms of the eigenvalues $E_\alpha$ and eigenfunctions $\vert \alpha \rangle$ of the single-particle problem as:
\[
\vert \psi (t) \rangle  = \frac{\sum_\alpha^\star \vert \alpha \rangle\langle \alpha \vert 0 \rangle \, e^{-i E_\alpha t / \hbar}}
{\sqrt{\sum_\alpha^\star \vert \langle \alpha \vert 0 \rangle \vert^2}}\, .
\]
In principle all eigenfunctions $\vert \alpha \rangle$ contribute to the sum. However in the many-body system, due to the scaling of the energies in the thermodynamic limit, the states that matters
physically, even the virtual ones, have all the same intensive energies (i.e., the states with an intensive energy different from zero will have a vanishing projection on $\vert 0 \rangle$). Since our single-particle toy model lacks, of course, this concentration property, we have to impose it as
an extra-constraint. The star above the sum in the previous equation means that we are restricting it to a subset of 
eigenstates of ${\cal H}$ which belongs to a small energy shell around the middle
of the band ($E_\alpha \in [-\Delta E, \Delta E]$),\cite{projector} and the denominator is a normalization factor that ensures that
$\vert \langle \psi (t) \vert \psi(t) \rangle \vert^2 = 1$ (see Ref.~[\onlinecite{PLMBL}] for a more extended discussion).

The wave-function amplitude in the infinite time limit on the $(k+1)k^{N-1}$ boundary site $x^{(N)}$ (when the time evolution is constrained only on the states close to zero energy) ${\cal W} (x^{(N)}) \equiv \lim_{t \to \infty} \left \vert \langle x^{(N)} \vert \psi (t) \rangle \right \vert^2$, has a very simple spectral representation in terms of the elements of the resolvent matrix
of the non-interacting Hamiltonian on the tree. 
The resolvent is defined as 
${\cal G}(z) = ({\cal H} - z {\cal I})^{-1}$, where ${\cal H}$ is given in Eq.~(\ref{eq:H}), 
${\cal I}$ is the identity matrix, and $z = E + i \eta$, $\eta$ being an infinitesimal imaginary regulator which smooths out the pole-like singularities 
in the denominator. As detailed in App.~\ref{app:spectral}, one finds: 
\begin{equation} \label{eq:amplitude}
\begin{aligned}
{\cal W} (x^{(N)}) &=
\frac{\sum_{\alpha}^\star \vert \langle x^{(N)} \vert \alpha \rangle \vert^2  \left \vert \langle \alpha \vert 0 \rangle \right \vert^2}
{\sum_\alpha^\star \vert \langle \alpha \vert 0 \rangle \vert^2} \\ 
& \approx \lim_{\eta \to 0^+} \frac{\eta \vert {\cal G}_{x^{(N)},0} (E=0) \vert^2}{{\rm Im} {\cal G}_{0,0} (E=0)} \, . 
\end{aligned}
\end{equation}
(To simplify the notations we will set $E = 0$ throughout unless specified differently.)
In the following, in the analytical study we shall consider $\eta = c \delta$ \cite{cvalue}
where  $\delta = 1/({\cal V} \rho)$ is the mean level spacing, which is the natural scale for the  imaginary regulator, 
and take the {\it simultaneous} limits ${\cal V} \to \infty$ and $\eta \to 0^+$.
Thanks to hierarchical structure of the lattice, the matrix elements of the resolvent on sites $0$ and $x^{(N)}$ can be explicitly 
written in terms of the diagonal elements of the
so-called ``cavity'' Green's functions $G_{x^{(i)} \to y^{(i-1)}}$ (see fig.~\ref{fig:chain}), which is 
defined as the diagonal element on site $x^{(i)}$ (belonging to the $i$-th generation of the tree) of the resolvent of a modified (``cavity'') Hamiltonian ${\cal H}_{x^{(i)} \leftrightarrow y^{(i-1)}}$ where the edge between sites $x^{(i)}$ and $y^{(i-1)}$ (belonging to the $(i-1)$-th generation of the tree) has been removed: $G_{x^{(i)} \to y^{(i-1)}} = ({\cal H}_{x^{(i)} \leftrightarrow y^{(i-1)}} - i \eta)_{x^{(i)},x^{(i)}}^{-1}$. 
 By progressively integrating out all the sites from the leaves to the root in the following expression or using matrix identities, one finds
(we set the hopping rate $t$ equal to $1$ throughout):
\begin{equation} \label{eq:corr}
\begin{aligned}
& {\cal G}_{x^{(N)},0}  = \frac{\int {\cal D} \varphi \, \varphi_{0} \, \varphi_{x^{(N)}} \, e^{-\frac{1}{2} \sum_{x,y} \varphi_x ( {\cal H} - i \eta)_{x,y} \varphi_y}}
{\int {\cal D} \varphi \, e^{-\frac{1}{2} \sum_{x,y} \varphi_x ( {\cal H} - i \eta)_{x,y} \varphi_y}} \\
& \,\,\,\,\,\,\, = G_{x^{(N)} \to x^{(N-1)}} G_{x^{(N-1)} \to x^{(N-2)}} \cdots G_{x^{(1)} \to 0} {\cal G}_{0,0} \, ,
\end{aligned}
\end{equation} 
Moreover, as shown in Ref.~[\onlinecite{ATA}], on tree-like structures the diagonal elements of such cavity Green's functions satisfy the following exact recursion relation:
\begin{equation} \label{eq:recursion}
G_{x \to y}^{-1} = \epsilon_x - i \eta - \!\! \sum_{z \in \partial x / y} \!\! G_{z \to y} \, ,
\end{equation}
where the sum runs over all the neighbors $z$ of $x$ except the cavity site $y$.
Hence, using Eqs.~(\ref{eq:amplitude}) and~(\ref{eq:corr}) ${\cal W} (x^{(N)})$ 
can be finally expressed as:
\[
 {\cal W} (x^{(N)}) \approx \frac{\eta \left \vert {\cal G}_{0, 0} \right \vert^2}{{\rm Im} {\cal G}_{0,0}} 
 \prod_{i=1}^N \left \vert G_{x^{(i)} \to x^{(i-1)}} \right \vert^2  \, .
\]
A measure of the delocalization, or ergodicity, of the quantum dynamics  
can then be obtained as the wave-function amplitude at the boundary of the tree $P_B^\infty$ (see App.~\ref{app:return}) obtained by summing the previous expression 
over all possible sites $x^{(N)}$ of the boundary of the tree (a proxy of the many-body configurations which are $O(N)$ spin flips away from the initial one):
\begin{equation} \label{eq:Pdecorr}
	P_B^\infty \equiv \sum_{x^{(N)}} {\cal W} (x^{(N)}) 
	\approx \frac{ \eta \left \vert {\cal G}_{0, 0} \right \vert^2}{{\rm Im} {\cal G}_{0,0}} \sum_{\textrm{\sf P}} \prod_{i=1}^N \left \vert G_{x^{(i)} \to x^{(i-1)}} \right \vert^2 \, ,
\end{equation}
where the sum in the numerator is over all directed paths ${\rm {\sf P}}$ of length $N$ connecting the leaves of the tree with the root through the edges $x^{(i)} \to x^{(i-1)}$ (one
of those paths is represented in fig.~\ref{fig:chain}). A $P_B^\infty$ different from zero in the large-$N$ limit is a signature that the system is delocalized, whereas on the contrary  $P_B^\infty=0$ indicates localization.  

Now let us highlight a fact that is central to our work: The sum (\ref{eq:Pdecorr}) is over an exponential
number of paths, $k^N$. 
In the large $N$ limit there are hence {\it two possible cases}: (1) The sum is dominated by few paths only; (2) The sum is dominated by an exponential number of paths $\overline{k}^N$ with an 
effective branching ratio $\overline{k}$ less than $k$ and disorder dependent. In the following we show that within the delocalized phase, $P_B^\infty>0$, there exists a sharp phase transition between these two regimes, and that such transition is related to the glass transition of directed polymer in random media.\cite{DPRM} 

Indeed, by introducing the edge-energies $\omega_{x^{(i)} \to x^{(i-1)}}$ and
the site-energy $\omega_{0}$ by setting $e^{-\omega_{x^{(i)} \to x^{(i-1)}}} = \vert G_{x^{(i)} \to x^{(i-1)}} \vert^2$ and $e^{- \omega_{0}} = \vert {\cal G}_{0, 0} \vert^2$, one can re-interpret the numerator of
Eq.~(\ref{eq:Pdecorr}) as the partition function of a directed polymer on the Cayley tree with $N$ generations in presence of quenched bond disorder (and quenched on-site disorder on the root) at inverse ``temperature'' $\beta=1$:\cite{gabriel}
\begin{equation}
\begin{aligned}
\sum_{x^{(N)}}  {\cal W} (x^{(N)}) 
&\approx \frac{\eta  e^{- \omega_{0}}}{{\rm Im} {\cal G}_{0,0}}  \sum_{\textrm{\sf P}} \prod_{i=1}^N
e^{-\omega_{x^{(i)} \to x^{(i-1)}}} \\
&= \frac{\eta Z_{\rm DP}(\beta=1,N)}{{\rm Im} {\cal G}_{0,0}}
\end{aligned} 
\end{equation}
By similar arguments, see App.~\ref{app:ZvsImG}, one can also relate the imaginary part of the resolvent ${\rm Im} {\cal G}_{0,0}$ to the partition functions $Z_{\rm DP} (\beta=1,M)$ of directed polymers of length $M$ starting at the root of the Cayley tree and ending at the $M$-th generation:
\begin{equation} \label{eq:ImGZpre}
        {\rm Im} {\cal G}_{0,0} = \sum_{M=0}^N \eta Z_{\rm DP} (\beta = 1,M) \, .
\end{equation}
In conclusion, the thermodynamic properties of the associated directed polymer problem are instrumental in studying 
the delocalization and ergodicity properties of Anderson localization.\cite{garelDP} In the following we will study them in detail. 

\subsection{Glass transition of directed polymer in random media}

In the original problem introduced by Derrida and Spohn the disordered consisted in i.i.d. onsite energies only. The DPRM can however be solved even in the case of correlated onsite and link disorder (the $| G_{x^{(i)} \to x^{(i-1)}} |^2$ are correlated along a path)\cite{noi,garel,DPRM,garelDP} as we now recall. \\
One has to compute the generalized average ``free-energy'' (also introduced in Refs.~[\onlinecite{levy}] and~[\onlinecite{aizenmann}]):
\begin{equation} \label{eq:DPRMphi}
	\phi(\beta) = \lim_{N \to \infty} \frac{1}{\beta N} \left \langle \ln \left( \left \vert {\cal G}_{0, 0} \right \vert^{2 \beta} \sum_{\textrm{\sf P}} \prod_{i=1}^N \left \vert G_{x^{(i)} \to x^{(i-1)}} \right \vert^{2 \beta} \right)
	\right \rangle \, ,
\end{equation}
where the average is performed over the quenched random energies $\epsilon_x$ of the non-interacting tight-binding toy model~(\ref{eq:H}), 
which, once the fixed point of Eqs.~(\ref{eq:recursion}) is found, 
yield the effective random energy landscape for the DP.
The average free-energy is a convex function of $\beta$ and has a one-step RSB freezing glass transition, akin to the one of
the Random Energy Model (REM):\cite{DPRM,REM,spinglass} by decreasing the ``temperature'' $1/\beta$ the generalized free-energy decreases
until the critical point $\beta_\star$, defined by $\partial \phi (\beta) / \partial \beta |_{\beta_\star} = 0$,
is reached; for $\beta > \beta_\star$ the DP freezes and its free-energy
remains constant: In this glass phase the
number of paths contributing to (\ref{eq:DPRMphi}) is not exponential
in $N$, but instead $O (1)$, implying that the DP can be found only on few specific disorder-dependent paths with probability of order $1$,
whereas for $\beta < \beta_\star$ there is an exponentially small probability of finding the polymer on an exponentially large number of conformations.

The physical reason for that goes as follows:\cite{garel,DPRM,noi,levy} 
Denoting $e^{-N \beta f}$ the contribution of a given path of length $N$,
one can rewrite the sum in Eq.~(\ref{eq:Pdecorr}) as an integral over all paths
giving a contribution characterized by a value of $f$
between $f$ and $f + {\rm d} \! f$ times the number of such paths. By
denoting the latter $\exp(N \Sigma (f))$, one ends up with the
expression:
\[
	Z_{\rm DP} (\beta) = 
        \int {\rm d} \! f \, e^{N [- \beta f + \Sigma(f)]} \, .
\]
The value of $f$ that dominates the integral for $N \to \infty$
depends on $\beta$.  For small enough $\beta$, one finds that the
saddle point value of $f_\star (\beta)$ is such that $\Sigma(f_\star)>0$.
In this regime an exponential number of paths,
$\overline{k}^{N}$ (with $\overline{k} = e^{\Sigma(f_\star)}$), contributes to the sum.
By increasing $\beta$, $\Sigma(f_\star)$ decreases until the value
$\beta_\star$ is reached.  At this point the generalized entropy
$\Sigma(f_\star(\beta_\star))$ vanishes. Hence the generalized average free-energy
$\phi(\beta)$ is related to the Legendre transform of $\Sigma (f)$:
\[
	\phi(\beta) = -f_\star + \frac{\Sigma(f_\star)}{\beta} \, ,
\] 
and $\phi^\prime(\beta) = - \Sigma(f(\beta))/\beta^2$, and allows
one to find out whether a finite ($\Sigma(f_\star) = 0$) or an exponential ($\Sigma(f_\star) > 0$) number
of paths contributes to the partition function $Z_{\rm DP} (\beta)$. 


Since the physical value of our proxy for decorrelation, Eq.~(\ref{eq:Pdecorr}), is obtained for $\beta=1$, what matters here is whether the freezing of the DP takes place
at $\beta_\star$ above or below $1$.
In order to compute $\phi (\beta)$, on each edge of the lattice, and for a given value of $\beta$,
we introduce the variable
\[
	y_{x^{(i)} \to x^{(i - 1)}} = \sum_{\rm {\sf P_{N-i}}} \, \prod_{j = i}^{N} | G_{x^{(j)} 
        \to x^{(j-1)}} |^{2 \beta} \, ,
\]
where ${\rm {\sf P_{N-i}}}$ are all the directed paths of length
$N - i$ connecting the site $x^{(i)}$ to the boundary of the tree, and
$x^{(j)} \to x^{(j-1)}$ are all the directed edges (including
$x^{(i)} \to x^{(i - 1)}$) belonging to the path.
It is straightforward to
derive the following exact recursion relation for $y_{x^{(i)} \to x^{(i - 1)}}$:
\begin{equation} \label{eq:DPRMy}
        y_{x^{(i)} \to x^{(i - 1)}} = |G_{x^{(i)} \to x^{(i - 1)}}|^{2 \beta} \!\! 
        \sum_{x^{(i+1)} \in \partial x^{(i)}} y_{x^{(i+1)} \to x^{(i)}} 
        \, ,
\end{equation}
where $|G_{x^{(i)} \to x^{(i - 1)}}|^{2\beta}$ can  be  computed  using  Eq.~(\ref{eq:recursion}).
Eqs.~(\ref{eq:recursion}) and~(\ref{eq:DPRMy}) naturally lead to an exact
functional equation for the joint probability distributions $W_{i}^{(\beta)} (G,y)$
at the $i$-th generation of the tree:
\[
\begin{split}
	& W_i^{(\beta)} (G,y) = \! \int \! \textrm{d}p(\epsilon) \prod_{\ell=1}^k \textrm{d} W^{(\beta)}_{i+1} 
	(G_\ell, y_\ell) \\
	& \qquad \,\,
	\times \delta \! \left [ G^{-1} \! + \epsilon + i \eta + \sum_{\ell=1}^k G_\ell \right]  
	\delta \! \left [ y - |G_\ell|^{2 \beta} \sum_{\ell=1}^k
	y_\ell \right] \, ,
\end{split}
\]
where $p(\epsilon) = (1/W) \theta (-W/2 \le \epsilon \le W/2)$.
Once the fixed point of these equation has been found, one can obtain the the joint probability of ${\cal G}_{0,0}$ 
and $Z_{\rm DP}$ at the root of the tree as:
\[
\begin{split}
	& W_0^{(\beta)} ({\cal G},Z) = \! \int \! \textrm{d}p(\epsilon) \prod_{\ell=1}^{k+1} \textrm{d} W^{(\beta)}_{1} 
        (G_\ell, y_\ell) \\
	& \qquad \,\, \times \delta \! \left [ G^{-1} \! + \epsilon + i \eta + \sum_{\ell=1}^{k+1} G_\ell \right] 
	\delta \! \left [ Z - |G_\ell|^{2 \beta} \sum_{\ell=1}^{k+1}
        y_\ell \right] \, .
\end{split}
\]
The equations above can be solved by iteration using a population dynamics algorithm with arbitrary numerical 
precision. After $N$ generations one can then compute $\phi (\beta,N)$ as the average value of the
logarithm of $Z$
over the distribution $W_{0}^{(\beta)} ({\cal G},Z)$, divided by
$\beta N$, as in Eq.~(\ref{eq:DPRMphi}): $\phi (\beta,N) = \langle \ln Z \rangle_{W_0}/(\beta N)$.

\subsection{Numerical results I: Glass transition of DPRM and delocalized non-ergodic phase}
Performing the limit $N \to \infty$ requires an extrapolation of the numerical results obtained at finite 
$N$.~\cite{noi}
In order to avoid such extrapolation, the position of $\beta_\star$ can be more conveniently found computing the logarithm of the average partition function instead of the average of the logarithm. This leads to the so-called annealed free-energy: 
\begin{equation} \label{eq:DPRMphiann}
        \begin{aligned}
        \phi_{\rm ann}
                (\beta) & = \lim_{N \to \infty} \frac{1}{\beta N} \ln \left \langle 
		\left \vert {\cal G}_{0,0} \right \vert^{2 \beta} \sum_{\textrm{\sf P}} \prod_{i=1}^N \left 
		\vert G_{x^{(i)} \to x^{(i-1)}} \right \vert^{2 \beta}
		\right \rangle \\
                & = \lim_{N \to \infty} \frac{\ln \langle Z \rangle_{W_0}}{\beta N} \, .
        \end{aligned}
\end{equation}
whereas the one in Eq.~(\ref{eq:DPRMphi}) is called quenched free-energy. As discussed in Ref.~[\onlinecite{aizenmann}] (see also Refs.~[\onlinecite{REM}] and~[\onlinecite{spinglass}]) the two free-energies coincide for $\beta\le \beta_*$. Hence, the annealed and quenched free-energies can be equivalently used to identify the value of $\beta_\star$, 
but the annealed one much less computationally demanding.\cite{REM,spinglass}
Therefore one can use the annealed one to obtain $\phi(\beta)$ for $\beta\le \beta_*$, and the value of $\beta_*$, and then impose that $\phi(\beta)=\phi(\beta_*)$ for $\beta> \beta_*$.

\begin{figure}
\includegraphics[width=0.48\textwidth]{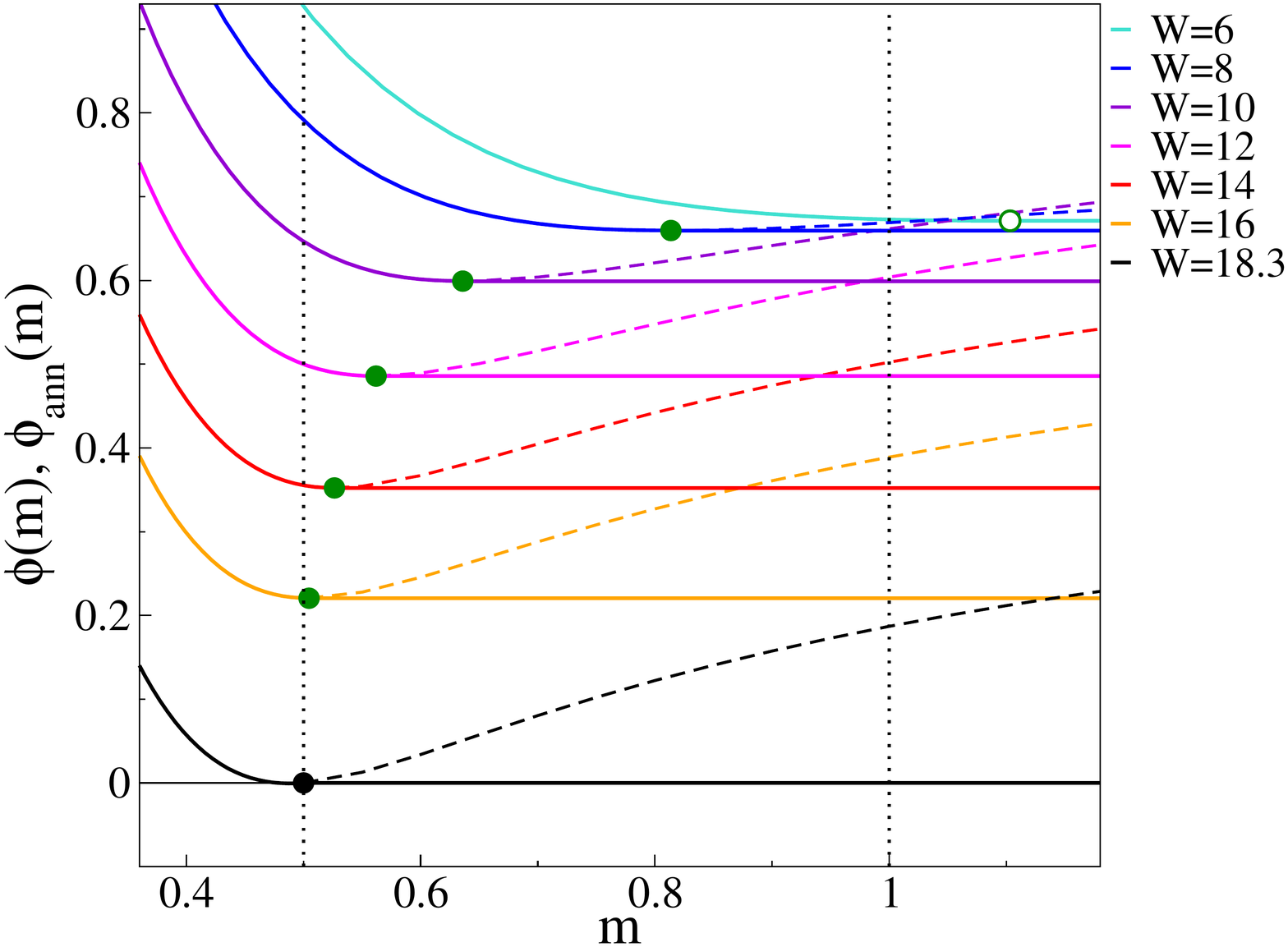}%
\caption{\label{fig:phi}
	Quenched (continuous lines) and annealed (dashed lines) average free-energy, $\phi (\beta)$ and $\phi_{\rm ann} (\beta)$, 
	Eqs.~(\ref{eq:DPRMphi}) and (\ref{eq:DPRMphiann}), as a function of the ``inverse temperature'' $\beta$
	of the DP on the Cayley tree associated to the delocalization of a particle at the root through the effective random energy landscape generated 
	by the $\vert G_{x_i \to x_{i-1}} \vert^{2}$'s, for different values of the disorder strength across the delocalized phase up to the localization transition.
	The 
	circles spot the position of the breaking point $\beta_\star$ where $\partial \phi (\beta) / \partial \beta |_{\beta_\star} = 0$, 
	which becomes greater than one for $W \lesssim W_T \approx 6.65$ and is equal to $1/2$ at $W_L \approx 18.3$ (where $\phi (1/2) = 0$).}
\end{figure}

We have obtained $\phi_{\rm ann} (\beta)$, for several values of the disorder across the whole delocalized side ($W \le W_L \approx 18.3$\cite{tikhonov_critical}) of the 
tight-binding Anderson model on the Cayley tree in the limit of large $N$, using the recursive 
equations~(\ref{eq:recursion}) and~(\ref{eq:DPRMy}).

\begin{figure}
\includegraphics[width=0.48\textwidth]{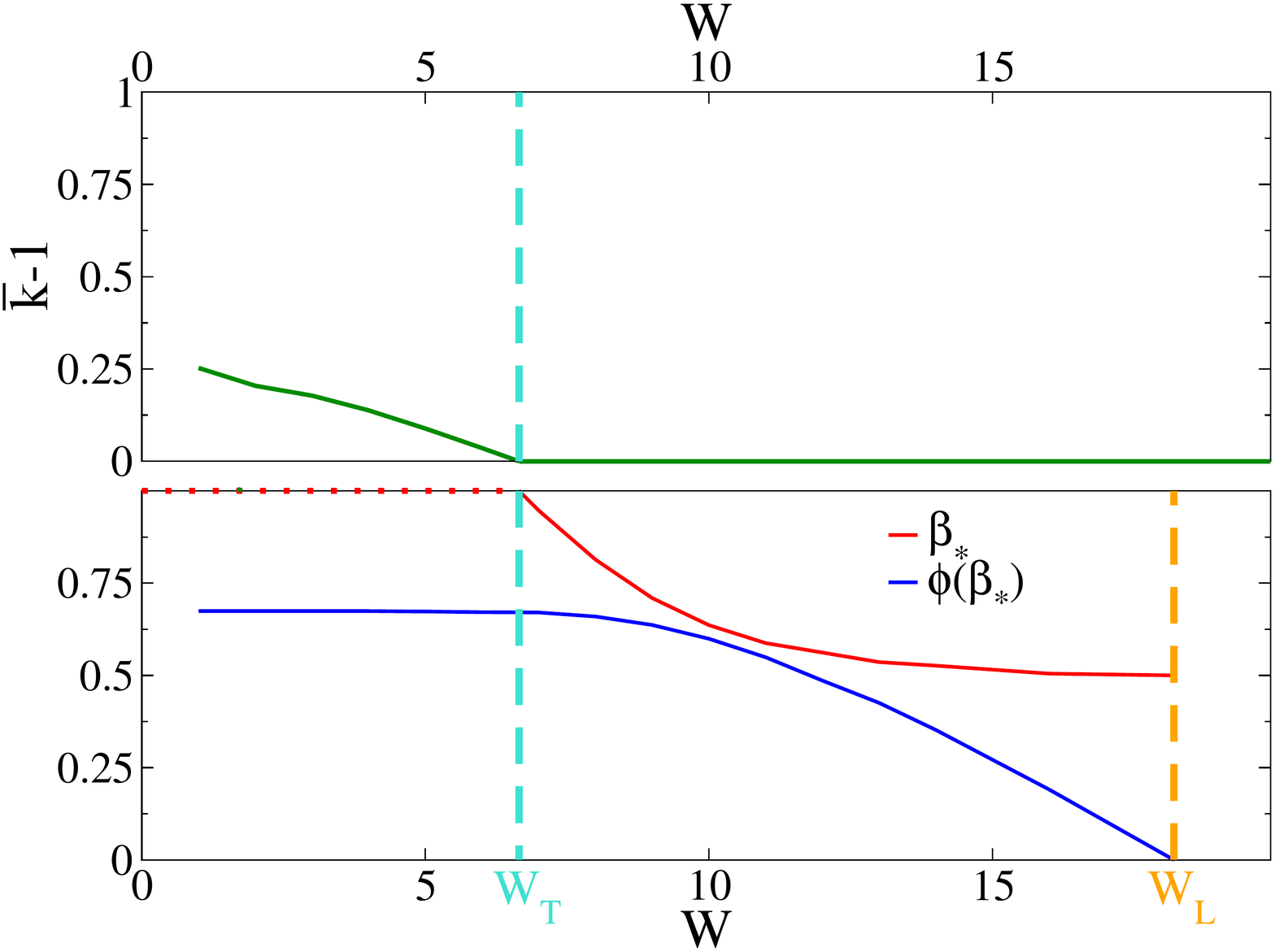}%
\caption{\label{fig:mstar}
Top panel: Effective branching ratio $\overline{k} = e^{- \phi^\prime(\beta=1)}$ (green) of the DP problem as a function of $W$. $\overline{k}$ vanishes at $W=W_T$ and remains zero in the glassy delocalized phase $W_T<W<W_L$.
Bottom panel:
        Critical ``inverse temperature'' $\beta_\star$ (red) and average free-energy $\phi(\beta_\star)$ (blue) as a function of $W$. At $W_T$ (turquoise dashed vertical line) $\beta_\star = 1$, while
	at $W_L$ (orange dashed vertical line) $\beta_\star = 0.5$ and $\phi(\beta_\star) = 0$.}
\end{figure}

We observe that $\phi(1)$ becomes positive below $W_L$. At this point, delocalization takes place and the wavefunction 
spreads far away from the root. 
As shown in App.~\ref{app:lyap}, $\phi(\beta = 1)$ coincides asymptotically with the Lyapunov exponent describing the growth ${\rm Im} {\cal G}$ under the iteration relations~(\ref{eq:recursion}),
which is a decreasing function of $W$
vanishing at $W_L$.\cite{ioffe1,ioffe3,ATA,aizenmann} 
We find that $\beta_\star \to 0.5$ and $\phi(\beta_\star) \to 0$ for $W \to W_L$. This 
is expected, as it was rigorously proved in Refs.~[\onlinecite{aizenmann}] and~[\onlinecite{aiz_war}], and indirectly found in Ref.~[\onlinecite{ATA}] (see also Refs.~[\onlinecite{garel}], [\onlinecite{noi}], and~[\onlinecite{levy}]). It is therefore a good check of our numerical method.  When diminishing
$W$ below $W_L$ the value of $\beta_\star$ increases and eventually reaches $1$ for
$W = W_T \approx 6.65$ (see Fig.~\ref{fig:mstar}), where the glass transition of the directed polymer takes place.
At weaker disorder, $W<W_T$, $\phi(\beta)$ is a smooth decreasing function of $\beta$ in the whole range $\beta \in [0.5,1]$. Hence, for $0< W < W_T$ the contribution to $\sum_{x^{(N)}} {\cal W} (x^{(N)})$ comes from an exponential number of paths,
while for $W_T < W < W_L$ delocalization occurs only on rare, ramified, specific paths which corresponds to the preferred disorder-dependent configurations of the polymer in the glassy phase.
In Fig.~\ref{fig:phi} we show plots of $\phi (\beta)$ for several values of $W$ across the delocalized phase, highlighting the position of $\beta_\star$.
Since the number of paths of length $N$ contributing to the sum in Eq.~(\ref{eq:Pdecorr}) scales as $e^{N\Sigma(f_\star)}$,  
the effective branching ratio $\overline{k}<k$ 
can be computed as $\overline{k} = e^{- \phi^\prime(\beta=1)}$.
In the top panel of Fig.~\ref{fig:mstar} we show that 
$\overline{k}$ associated to the exponential growth of the number of paths vanishes at $W=W_T$ and remains zero in the glassy delocalized phase $W_T<W<W_L$.\cite{exponential} In the bottom panel of Fig.~\ref{fig:mstar} we show the behavior of the critical ``inverse temperature'' $\beta_\star$ and of the average free-energy $\phi(\beta_\star)$ of the associated DP problem as a function of $W$.

\subsection{Numerical results II: Glass transition of DPRM, the singular statistics of the local density of states and multifractality} \label{sec:ldos}

\begin{figure}
\includegraphics[width=0.5\textwidth]{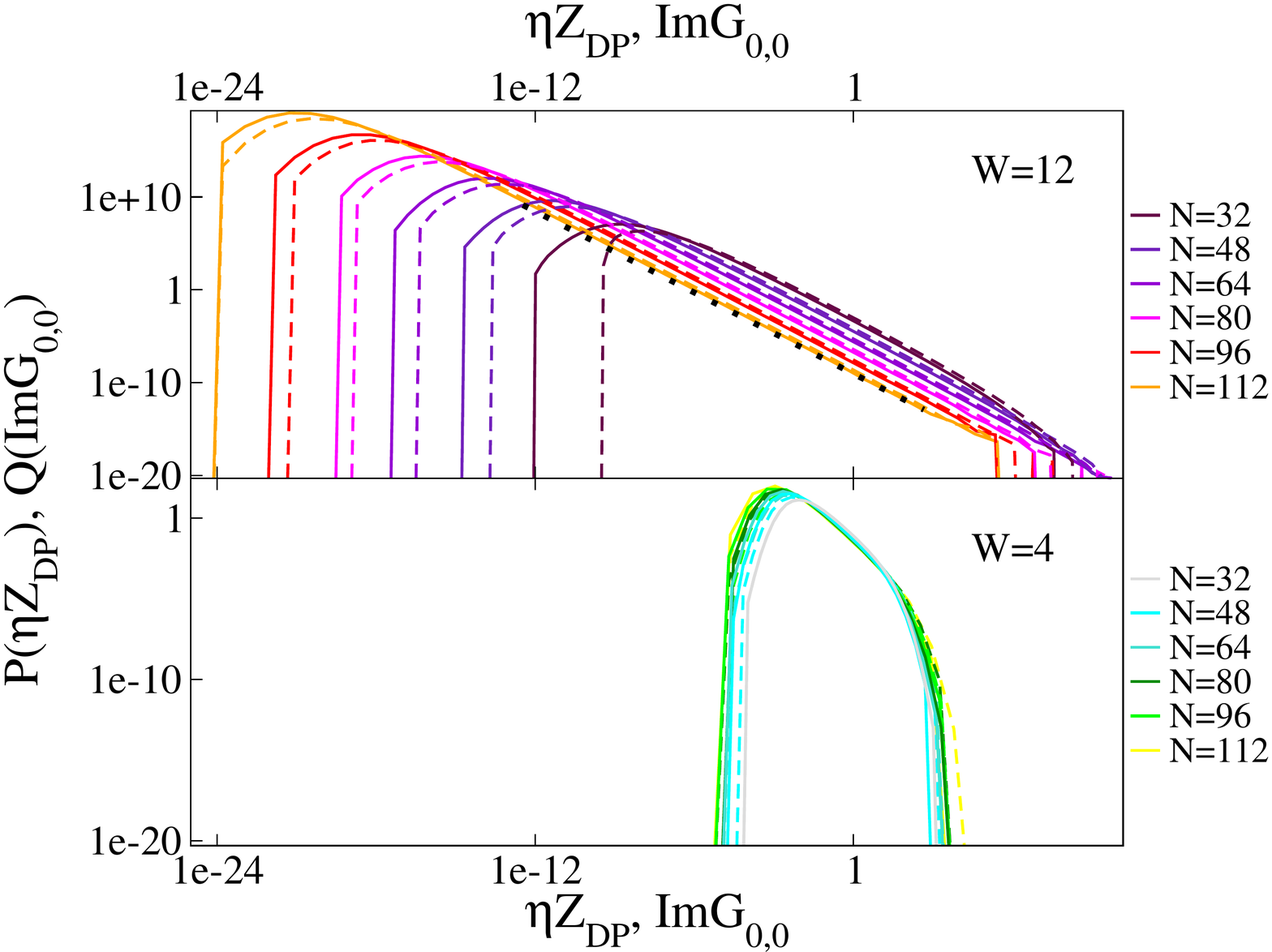}%
\caption{\label{fig:distrib}
	Log-log plot of the probability distributions $P(\eta Z_{\rm DP})$ (continuous lines) and $Q ({\rm Im} {\cal G}_{0,0})$ (dashed lines)
	for Cayley trees of $N$ generations (with $N$ going from $32$ to $112$), and for $W = 12 > W_T$ (top panel) and $W = 4 < W_T$ (bottom panel).
	The dotted black straight line shows the exponent $1 + \beta_\star \approx 1.54$ of the power-law tails of the distributions.}
\end{figure}

It was shown in Ref.~[\onlinecite{DPRM}] (see also Ref.~[\onlinecite{garelDP}]) that in the freezing glass
phase the DP partition function is a power-law tailed distributed random variable with an exponent $1+\beta_\star$
(and possibly logarithmic corrections). 
In fig.~\ref{fig:distrib} we confirm this result in our case: we show the probability distribution $P(\eta Z_{\rm DP})$ (we will omit to specify that $Z_{\rm DP}$ is computed at $\beta=1$ henceforth to simplify the notation) 
for Cayley tree of different sizes ($N=32, \ldots, 112$) , 
for $W=12$ (top panel) and
$W=4$ (bottom panel), showing that the statistics of $\eta Z_{\rm DP}$ 
is completely different below and above $W_T$.
For $W<W_T$, $P(\eta Z_{\rm DP})$ 
converges to a (size-independent) stable non-singular distribution which decreases fast 
to zero at large arguments.
For $W>W_T$, instead, $P(\eta Z_{\rm DP})$ 
has a singular behavior in the limit $N \to \infty$ (and $\eta \to 0$):
the typical value of $\eta Z_{\rm DP}$ 
goes to zero in the large $N$ limit, while 
its average stays finite, and is dominated by the fat tails of the PDF which are characterized by an exponent $1+\beta_\star$ (the power-law behavior is cut-off for
$\eta Z_{\rm DP} \sim 1/\eta$).

This result has important implications on the distribution of the local density of states ${\rm Im} {\cal G}_{0,0}$. 
In fact, as discussed in App.~\ref{app:ZvsImG}, ${\rm Im} {\cal G}_{0,0}$ is directly connected to the partition functions of the DPs 
via Eq.~(\ref{eq:ImGZpre}). 
Since in the delocalized phase, $W<W_L$, $Z_{\rm DP} (\beta = 1,M)$ grows with $M$ one expect the sum 
in~(\ref{eq:ImGZpre}) to be dominated by the last term, i.e. that the main contribution to ${\rm Im} {\cal G}_{0,0}$
is given by $Z_{\rm DP} (\beta = 1,N)$. This in turn implies that the singular statistics found for $Z_{\rm DP} (\beta = 1,N)$
 in the delocalized glassy phase also holds for ${\rm Im} {\cal G}_{0,0}$. 

In Fig.~\ref{fig:distrib} we show that this is indeed the case: we plot the probability distributions $Q ({\rm Im} {\cal G}_{0,0})$ of the imaginary part of the Green's functions at the root of the tree,
which as expected coincides essentially with $P(\eta Z_{\rm DP})$. 

This singular statistics of the local density of states implies multi-fractal behavior for $W_T<W<W_L$. In fact, since $\beta_\star \le 1$, the tails of $Q ({\rm Im} {\cal G}_{0,0})$ give the leading contribution to the DoS and to all the moments
$\langle ({\rm Im} {\cal G}_{0,0})^q \rangle$ with $q>\beta_\star$, whereas the bulk part only yields a vanishing one.
Conversely, all the moments $\langle ({\rm Im} {\cal G}_{0,0})^q \rangle$ with $q<\beta_\star$ are dominated by the behavior
of the typical value. (The probability distribution of the real part of ${\cal G}_{0,0}$ instead converges for
$N \to \infty$ to a stationary non-singular distribution.)
Since the moments of ${\rm Im} {\cal G}_{0,0}$ are related to the moments of the wave-functions'
amplitudes at the root of the Cayley tree,
for $W > W_T$ one has a bifractality scenario in the vicinity of the root 
which
is exactly the same as the one recently found in Ref.~[\onlinecite{mirlinCT}] using the
supersymmetric non-linear $\sigma$-model approach for the $p$-orbital Anderson model on the Cayley tree with
$p \gg 1$.
Remarkably enough, the solution of the problem in that case is found in terms of the Fisher-KPP equation, which was
first introduced in Ref.~[\onlinecite{KPP}], and is known to emerge in a broad class of non-linear problems describing
propagation of a front between an unstable and a stable phase, including DPRM.\cite{DPRM,derrida}
The transition discussed here is also of the same kind as the one found in Ref.~[\onlinecite{ioffe3}] by mapping the
 iteration equation for the imaginary part of the Green's function on the traveling wave problem,\cite{DPRM}
and using a RSB formalism for a slightly modified distribution of the on-site random energies and in
the large connectivity limit $k \to \infty$.

\section{Quantum dynamics on the Cayley tree and the depinning tansition of the directed polymers} 
\label{sec:dynamics}
In this section we follow the ideas of  Ref.~[\onlinecite{PLMBL}] and study the relaxation of proxies of average and correlation functions of local operators in real space on the delocalized side of the phase diagram, 
by using the Anderson model on the Cayley tree~(\ref{eq:H}) as a toy model for the many-body dynamics. 
The observables we focus on are the counterpart of the imbalance and of the (infinite temperature) equilibrium correlation function.
The imbalance measures whether an initial random magnetization profile converges to its flat thermodynamic average or remains
instead inhomogeneous even at very long time, keeping memory of the initial configuration.
For a $N$-body disordered isolated quantum system, described for instance by Eq.~(\ref{eq:HMB}), this corresponds to check whether
$(1/N) \sum_i \langle \sigma_i^z (t) \rangle^2$ tends to zero or to a positive residual value at long times.
Within our analogy between MBL and single-particle Anderson localizaiton in a high-dimensional space, the counterpart of a random initial
state corresponds to a wave-function at $t=0$ localized on a particular site $x_0$ of the lattice (that we will take as the root of the tree)
with energy close to zero: $\vert \psi (t=0) \rangle = \vert 0 \rangle$, such that $\epsilon_{0} \approx 0$.
In order to study averages and correlations of spin operators in real space, we need to define Bethe lattice proxies of such local operators.
The representation of $\hat{\sigma}_i^z$ in Fock space is simply $\hat{\sigma}_i^z = \sum_{\{ \sigma_i^z \}} \vert \{ \sigma_i^z \} \rangle \langle \{ \sigma_i^z \}
\vert f(\{ \sigma_i^z \})$, where $f(\{ \sigma_i^z \})$ is equal to the value of $\sigma_i^z$ in the configuration $\{ \sigma_i^z \}$.
The properties of this function is that it varies in a rapid and scattered way along the hyper-cube and is equal to $+1$ or $-1$ on half
of the configurations. Here, as done in Ref.~[\onlinecite{PLMBL}], we approximate such complex behavior by a random one on the Bethe lattice, by
defining a proxy $\hat{O}$ of the local operator $\hat{\sigma}_i^z$ as:
\begin{equation} \label{eq:Oproxy}
\hat{O} \equiv \sum_x \vert x \rangle \langle x \vert f(x) \, ,
\end{equation}
where $f(x)$ is a i.i.d. random
binary variable equal to $\pm 1$ with probability $1/2$. By doing so, we only keep the statistical properties of the coefficients $f(x)$ but neglect its correlations and its specific structure. Using this definition we have that:
\[
	\begin{aligned}
		\langle O(t) \rangle &= \langle 0 \vert e^{i H t / \hbar} \hat{O} \, e^{-i H t/\hbar} \vert 0 \rangle \\ & = \sum_{x=1}^{\cal V} f(x) \left \vert
	\sum_\alpha^\star \langle x \vert \alpha \rangle \langle \alpha \vert 0 \rangle e^{-i E_\alpha t/\hbar } \right \vert^2 \, ,
	\end{aligned}
\]
where, as explained above, the sum over the eigenstates of the single-particle Hamiltonian~(\ref{eq:H}) is restricted only
on the eigenvectors with energies $E_\alpha$ within a small bandwidth $[-\Delta E, \Delta E]$ around zero energy, in order to mimic the fact that in a many-body
systems the only states that contribute to the time evolution have all the same intensive energy.~\cite{projector} In a $N$-body system this
restriction is automatically enforced by the scaling of the energies in the thermodynamic limit, but our toy model~(\ref{eq:H}) lacks 
of this concentration properties and we need to enforce it by hand.

Averaging over the random variables $f(x)$ and on the random on-site energies $\epsilon_x$ we obtain the Bethe lattice proxy of the Imbalance:
\begin{equation}
	I(t) = \frac{\sum_{x=1}^{\cal V} \left \vert
	\sum_\alpha^\star \langle x \vert \alpha \rangle \langle \alpha \vert 0 \rangle e^{-i E_\alpha t/\hbar } \right \vert^4}
	{\sum_{x=1}^{\cal V} \left \vert
	\sum_\alpha^\star \langle x \vert \alpha \rangle \langle \alpha \vert 0 \rangle \right \vert^4}\, .
\end{equation}
Note that because of the constraint of the sum over the
eigenstates the numerator is not equal to one for $t=0$ and we cure this pathology by normalizing $I(t)$ by
its value at $t=0$.

Following the same kind of reasoning one can define the Bethe lattice proxy of the (infinite temperature) equilibrium correlation function:
\begin{equation} \label{eq:Ct}
	\begin{aligned}
		C(t) &= \overline{\langle (O (t) O(0) + O(0) O(t)) \rangle} \\ &= \frac{\sum_{\alpha,\beta}^\star \left \vert \langle 0 \vert \alpha \rangle \right \vert^2
	\left \vert \langle 0 \vert \beta \rangle \right \vert^2 \cos \left [ \left (E_\alpha - E_\beta \right) t/\hbar \right ]}
		{\left (\sum_{\alpha}^\star \left \vert \langle 0 \vert \alpha \rangle \right \vert^2 \right )^2}\, ,
	\end{aligned}
\end{equation}
where the average is performed over the random coefficients $f(x)$ of the local operator $\hat{O}$ and the random on-site energies $\epsilon_x$ of the Anderson tight-binding toy model.
Note that $C(t)$ actually coincides with the so-called return probability, which will be more extensively discussed in App.~\ref{app:return} and whose 
time dependence on the RRG has been recently analyzed in Refs.~[\onlinecite{scardicchio_dyn}] and~[\onlinecite{mirlintikhonov}].

The time evolution of $I(t)$ and $C(t)$ has been studied on the RRG in Ref.~[\onlinecite{PLMBL}], where we showed that at moderately large time they both 
display unusually slow relaxations and power-law-like behaviors strikingly similar (at least visually and on moderately large time-scales) to the ones observed  in
recent experiments and simulations in the bad metal phase of many-body disordered isolated quantum systems approaching the
MBL transition.\cite{dave1,BarLev,demler,alet,torres,luitz_barlev,doggen,evers,experiments1,experiments2,experiments3} 
This occurs in a broad range 
of disorder where previous studies have suggested that the eigenfunctions of the Anderson model on the RRG might be delocalized but 
non-ergodic.\cite{noi,ioffe1,ioffe3} 
More recently, it was shown that for larger system sizes and for larger times the apparent algebraic decay is in fact cut-off and
replaced by an exponential one.\cite{scardicchio_dyn,mirlintikhonov}

\begin{figure}
\includegraphics[width=0.5\textwidth]{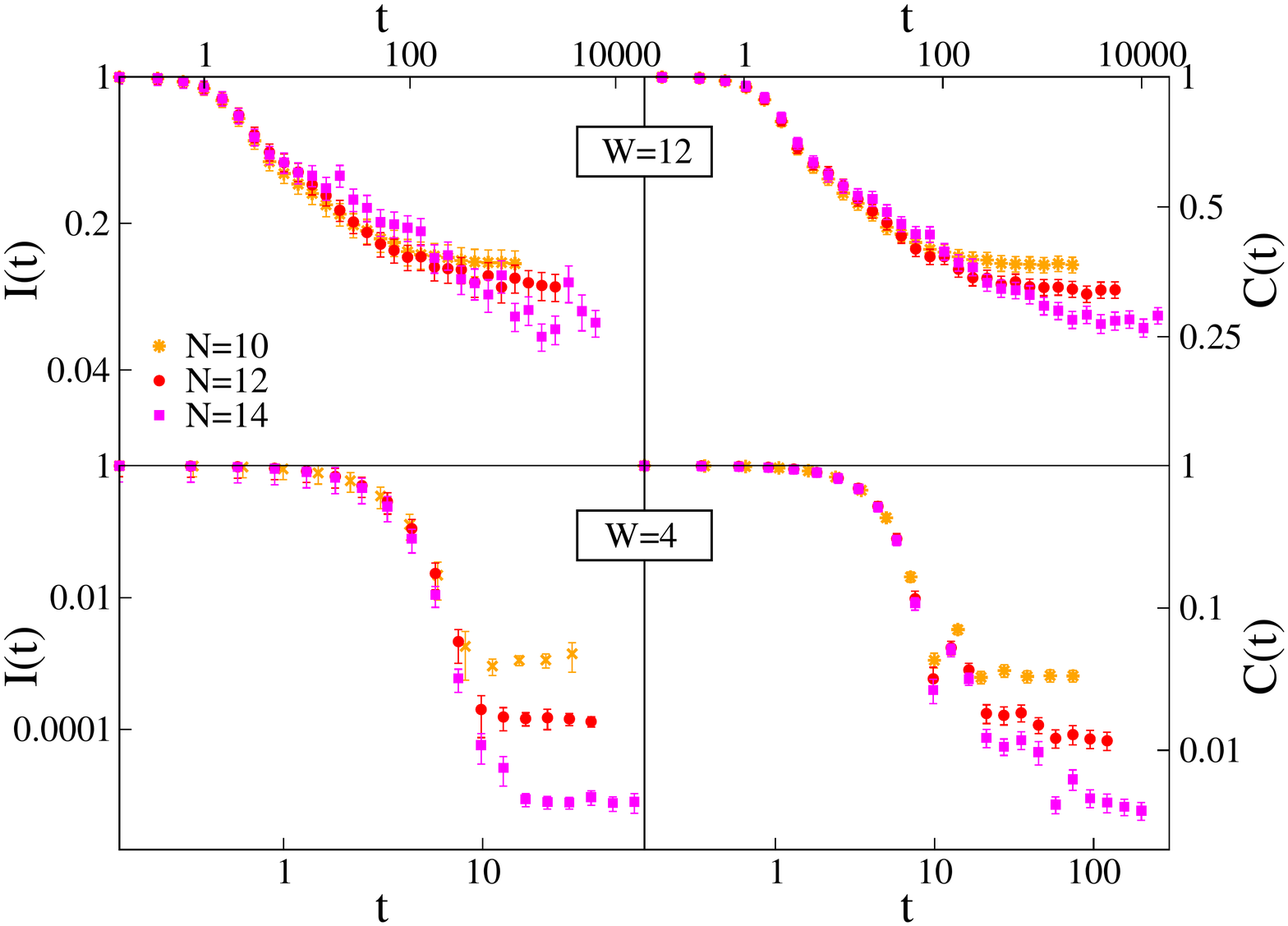}%
\caption{\label{fig:dyn}
	$I(t)$ (left panels) and $C(t)$ (right panels) as a function of $t$ for Cayley tree of different sizes and for
	$W=12$ (top panels) and $W=4$ (bottom panels).}
\end{figure}

Here we repeat the analysis of Ref.~[\onlinecite{PLMBL}] using the Cayley tree as the underlying lattice mimicking the Fock space instead of the RRG, and we will come back on the effect of the loops in Sec.~\ref{sec:loops}.
The dynamical evolution of the imbalance and of the equilibrium dynamical correlation function are plotted in fig.~\ref{fig:dyn} for 
different system sizes and for two values of the disorder strength, $W=4$, below the DP glass transition, and $W=12$, above the DP glass
transition, but still in the delocalized phase (the data are averaged over $2^{21-N}$ samples).
The figures clearly show that for $W>W_T$ a regime of slow dynamics sets in where both observables exhibit a power-law behavior 
which extends to larger and larger times as the system size is increased, whereas for $W<W_T$ the relaxation is fast and exponential (the plateau observed at large times is a finite size effect and goes to zero in the thermodynamic limit).

\subsection{Depinning transition of the Directed Polymers and relationship with the power-law relaxation} \label{sec:avalanches}

Such behavior of the quantum dynamics on the Cayley tree indicates that the emergence of the slow and unusual relaxation can not be simply related to the non-ergodicity of the wave-functions. 
In fact, all eigenstates of the Anderson model on the Cayley tree are multifractal in the whole delocalized phase, even below $W_T$,\cite{garelDP,garel,mirlinCT,Bethe} yet $I(t)$ and $C(t)$ exhibit a fast exponential decay in a region $0<W<W_T$ (see the bottom panels of fig.~\ref{fig:dyn}). The slow dynamics only sets in for $W > W_T$ and thus must be tightly related to the freezing glass transition of the paths along which the wave-function spreads. 
More formally, a direct link can be established between the depinning transition of the DP in the glass phase\cite{gabriel} (e.g., when a parameters like the energy is varied) and the the singular behavior of the ({\it local}) overlap correlation function, defined as follows: 
\[
K_2^{(0)} (E) =  
\left \langle
{\cal V}^2 \sum_{\alpha,\beta}^\star 
| \langle 0 \vert \alpha \rangle |^2 | \langle 0 \vert \beta \rangle |^2 \delta 
        \big[ E - (E_{\alpha} - E_{\beta}) \big]
        \right \rangle \, ,
\]
where $\langle 0 \vert \alpha \rangle$ is the amplitude of the eigenvector $\vert \alpha \rangle$ on the root of the tree.

The interest of introducing such spectral probe 
is twofold.
On the one hand, this function displays different scaling behaviors for the ergodic, localized, and multifractal states, and can be thus used to 
probe  the  non-ergodic  delocalized phase:\cite{ioffe3,Bethe,kravtsov}
For eigenfunctions of GOE matrices $K_2^{(0)} (E) = 1$ identically, independently on $E$ on the entire spectral band-width.
In the standard (ergodic) metallic phase 
$K_2^{(0)} (E)$ has a plateau at small energies ($E < E_{\rm Th}$), followed by
a fast-decay which is described by a power-law, $K_2^{(0)} (E) \sim E^{-\gamma}$, with a system-dependent exponent.\cite{chalker_K2} 
The height of the plateau is larger than one, which implies an enhancement of correlations compared to the
case of independently fluctuating Gaussian wave-functions.
The Thouless energy, $E_{\rm Th}$, which separates the 
plateau from the power-law decay, stays finite in the thermodynamic limit and extends to larger and larger energies as one goes deeply
into the metallic phase, and corresponds to the energy range over which GOE-like correlations establish.\cite{thouless} 
The (expected) behavior of the overlap correlation function for multifractal eigenfunctions is instead drastically different:\cite{kravtsov} 
The plateau is present only in a narrow energy interval $E < E_{\rm Th} \sim {\cal V}^{D_2-1}$
which shrinks to zero in the thermodynamic limit, while its height grows ${\cal V}^{1-D_2}$.
This can be interpreted recalling that multifractal wave-functions typically occupy a fraction ${\cal V}^{D_2}$ of the total sites,
which implies the existence of an energy scale, $E_{\rm Th}$, which decreases with ${\cal V}$ but stays much larger than the
mean level spacing, beyond which eigenfunctions poorly overlap with each other and the statistics is no longer GOE.

On the other hand, $K_2^{(0)} (E)$ is essentially the Fourier transform of our
proxy for the equilibrium correlation function (i.e., the return probability), Eq.~(\ref{eq:Ct}):\cite{DPRM,mirlintikhonov}
\begin{equation} \label{eq:CtFT}
C(t) \approx C_\infty + \frac{ 
\int_\delta^{2 \Delta E} \de E \, K_2^{(0)} (E) \, \cos(Et/\hbar)} 
{\left (\sum_{\alpha}^\star \left \vert \langle 0 \vert \alpha \rangle \right \vert^2 \right )^2}
\, .
\end{equation}
The behavior at large times of $C(t)$ is thus tightly related to the behavior at small $E$ of $K_2^{(0)} (E)$, i.e., the power-laws observed in the time decay of $C(t)$ are linked to the power-laws in energy found for $K_2^{(0)} (E)$.
(The statistics of the infinite time limit of the return probability, $C_\infty$, will be discussed separately in App.~\ref{app:return}.)
For any given random instance of the Hamiltonian~(\ref{eq:H}),
the overlap correlation function at the root of the tree~(\ref{eq:corr}) can be easily expressed in terms of the Green's functions on site $0$
computed at energies $\pm E/2$.\cite{ioffe3,Bethe,mirlintikhonov} Since, as discussed above and in App.~\ref{app:ZvsImG}, the imaginary part of the Green's function at the root of the tree 
coincides essentially with ($\eta$ times) the partition function of the DP, one finally finds:
\begin{widetext}
\begin{equation} \label{eq:K2}
\begin{aligned}
        K_2^{(0)} (E) &= \! \lim_{\eta \to 0^+} \! \left \langle  \frac{{\cal V}^2 \, {\rm Im}{\cal G}_{0,0} (-E/2) \, {\rm Im}{\cal G}_{0,0} (E/2)}
        {\sum_{x=1}^{\cal V} {\rm Im}{\cal G}_{x,x} (-E/2) \sum_{x=1}^{\cal V} {\rm Im}{\cal G}_{x,x} (E/2)} \right \rangle 
        \approx \lim_{\eta \to 0^+} \left( \pi \rho \right)^{-2}\left \langle \eta Z_{\rm DP} (-E/2) \, \eta Z_{\rm DP} (E/2) \right \rangle \, .
\end{aligned}
\end{equation}
Thus, the equilibrium dynamical correlation function $C(t)$, defined in Eq.~(\ref{eq:Ct}) and studied in the previous section, is essentially the Fourier transform of the correlation function of $Z_{\rm DP}$ computed at two different values of the energy for the same disorder realization (and then averaged over the disorder).
As discussed previously, it is the behavior at small energy of $K_2^{(0)} (E)$ that plays a key role in determining the correlation function at long-times. A GOE-like trend for which  $K_2^{(0)} (0)$ is finite leads to a fast decorrelation in time, whereas a power law with negative exponent and, hence, a divergent $K_2^{(0)} (0)$, is associated to effective power laws for $C(t)$. By using our result that the distribution of  $\eta Z_{\rm DP} (0)$ is well behaved and has a finite second moment  
for $W<W_T$, whereas it is a power law with a diverging second moment for $W_T<W<W_L$, we can therefore directly link 
the anomalously slow dynamics in the bad metal phase to the glassy regime found for $W_T<W<W_L$.
\end{widetext}

\begin{figure}
\includegraphics[width=0.48\textwidth]{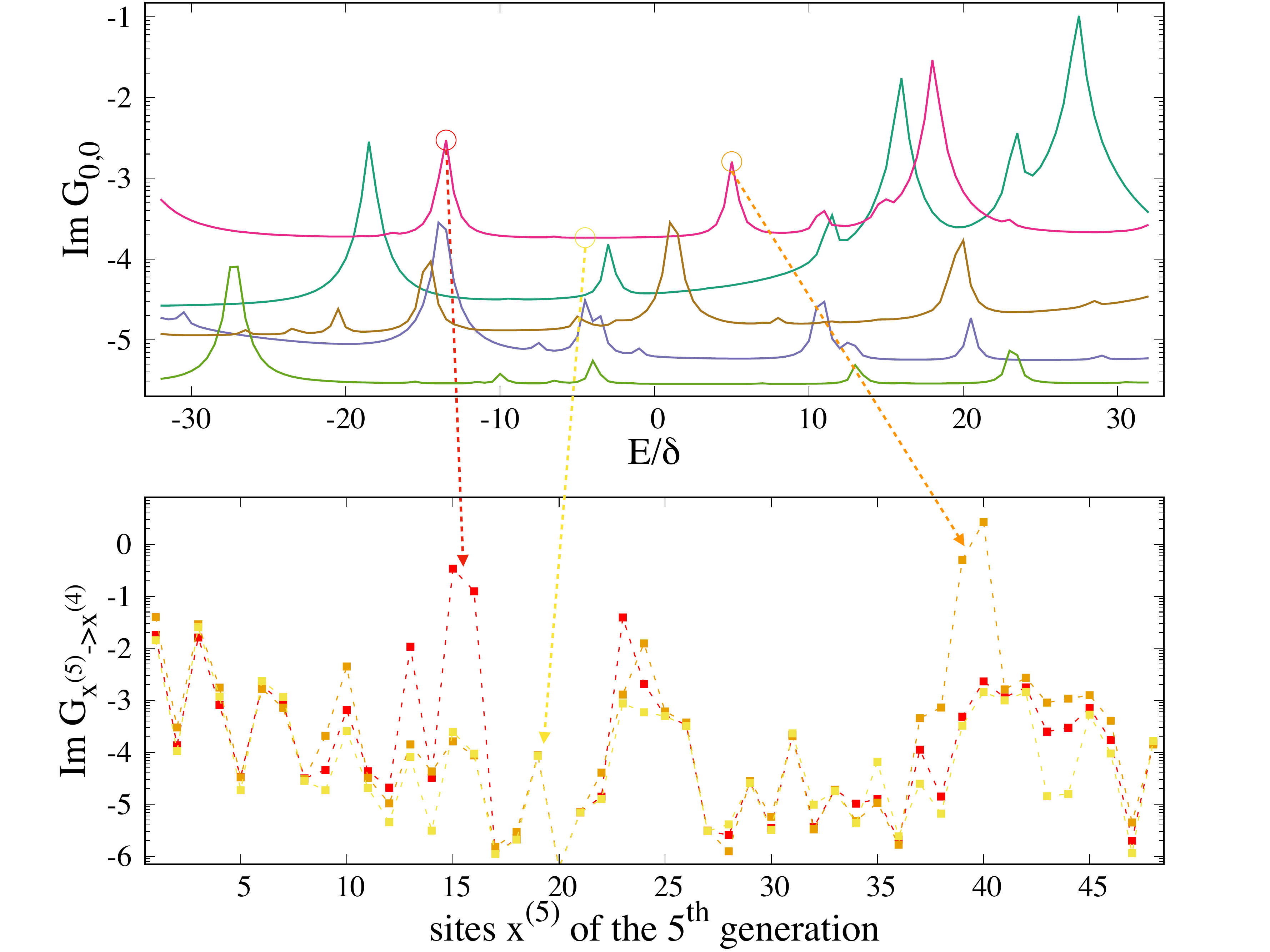}%
\caption{\label{fig:avalanches}
        Top panel: ($\log_{10}$ of the) Imaginary part of the Green's function ${\rm Im}{\cal G}_{0,0}$ at the root of a Cayley tree of $N=16$ generations as a function of the energy $E$ (measured in units of the mean level spacing $\delta$) at disorder $W=12>W_T$ for five independent realizations of the disorder. ${\rm Im}{\cal G}_{0,0}$ is pinned most of the time but jumps abruptly to very different values at some specific values of $E$ which depend on the specific disorder realization.
Bottom panel: ($\log_{10}$ of the) Imaginary part of the (cavity) Green's function ${\rm Im}{\cal G}_{x^{(5)} \to x^{(4)}}$ on the sites $x^{(5)} = 1, \ldots , 48$ of the $5$-th generation of the tree for three specific values of the energy spotted by the red, orange, and yellow circles in the top panel (and for a the disorder realization which corresponds to the magenta curve of the top panel).The difference between the two configurations of ${\rm Im}{\cal G}_{x^{(5)} \to x^{(4)}}$ is very small on most of the sites and of $O(1)$ on few sites only. This corresponds to a macroscopic jump between two preferred paths of the DP.  }
\end{figure}

A more physical insight can be gained analyzing by studying the behavior of the imaginary part of the Green's function. We start by plotting
in the top panel of fig.~\ref{fig:avalanches} the variation of ${\rm Im}{\cal G}_{0,0}$ at the root of a Cayley (of $N=16$ generations) when the energy is continuously varied on the scale of the mean level spacing $\delta$ close to the band center for four independent realization of the disorder.
We notice that ${\rm Im}{\cal G}_{0,0}$ is roughly constant (and small) over broad energy intervals (typically much larger than $\delta$) and has abrupt spikes around some specific values of $E$. 
Such spikes correspond to the existence of preferred conformations of the DP giving a large contribution to the partition function  $Z_{\rm DP}$  (and hence to ${\rm Im}{\cal G}_{0,0}$) at that particular value of the energy. As the energy is varied, the polymer is pulled away from such preferred conformation until a new one is found. (A similar behavior has been recently found in $2d$ Anderson localization in the strong disorder regime.\cite{gabriel}) This is shown in the bottom panel of fig.~\ref{fig:avalanches}, where we plot the imaginary part of the cavity Green's function for two specific values of the energy for which a maximum of ${\rm Im}{\cal G}_{0,0}$ is found (for a given disorder realization), focusing on the sites $x^{(5)} = 1, \ldots , {\cal V}_5$
of the $5$-th generation of the tree (where ${\cal V}_i = 3 \cdot 2^{i-1}$ is the total number of sites belonging to the $i$-th generation of the tree). ${\rm Im}{\cal G}_{x^{(5)} \to x^{(4)}}$ is pinned (and small) on most of the sites, and differs only on few points between the two configurations. This is the manifestation of the role played by rare events in determining ${\rm Im}{\cal G}_{0,0}$, and the depinning transition of the DP when the energy is varied, resulting in a macroscopic jump (i.e., an ``avalanche'') between two preferred directed paths: The preferred path which contribute the most to the first spike of ${\rm Im}{\cal G}_{0,0}$ (i.e., of the partition function $Z_{\rm DP}$) passes mainly through $x^{(5)} = 15$ and $x^{(5)} = 16$, while the preferred path contributing to the second spike passes mainly through $x^{(5)} = 39$ and $x^{(5)} = 40$.
For $W<W_T$, instead, ${\rm Im}{\cal G}_{0,0}$ is a smooth function of $E$.

We suspect that is the existence of such depinning transition of the DP \cite{yoshino}, which results in the macroscopic rearrangements of the conformation of the directed paths, which contribute the most to the sum of Eq.~(\ref{eq:Pdecorr}), produces a singular behavior (in the thermodynamic limit) of $\langle \eta Z_{\rm DP} (-E/2) \, \eta Z_{\rm DP} (E/2) \rangle$ 
and of the overlap correlation function $K_2^{(0)} (E)$, Eq.~(\ref{eq:K2}), at small energy difference, $E \sim \delta$.  This is also a direct manifestation of the fact that wave-functions closeby in energy display anomalously large correlations: In the glassy phase of the DP problem, $W>W_T$, eigenfunctions whose energy distance is of the order of few level spacing occupy typically the same paths on the tree, while eigenstates whose energy separation is larger than the typical distance between two spikes of fig.~\ref{fig:avalanches} poorly overlap.

\begin{figure}
\includegraphics[width=0.48\textwidth]{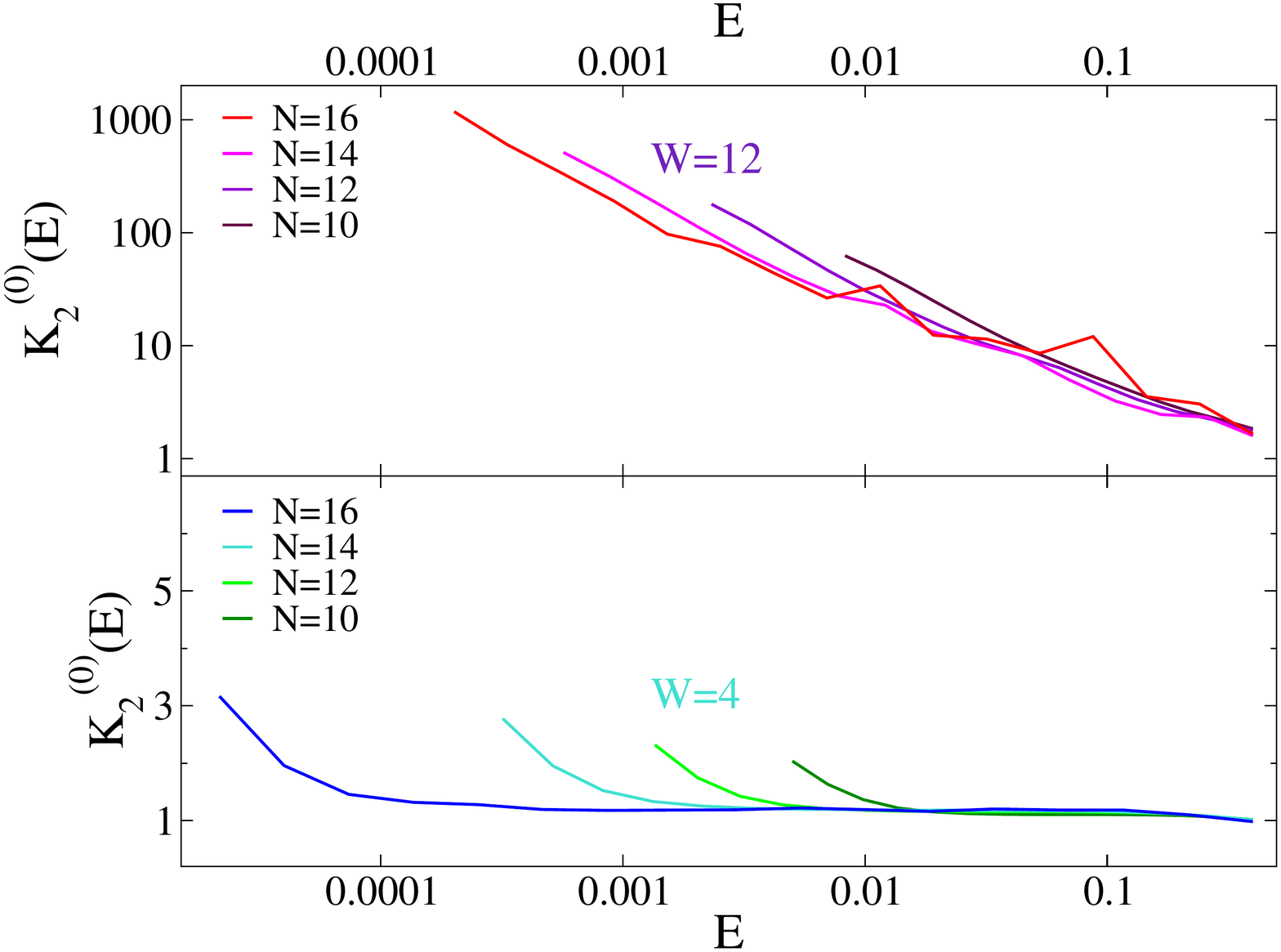}%
\caption{\label{fig:K2}
        Overlap correlation function $K_2^{(0)} (E)$ as a function of the energy for Cayley trees of several system sizes ($N=10$, $12$, $14$, and $16$ generations) and for $W=12>W_T$ (top panel) and $W=4<W_T$ (bottom panel).}
\end{figure}

\begin{figure}
\includegraphics[width=0.48\textwidth]{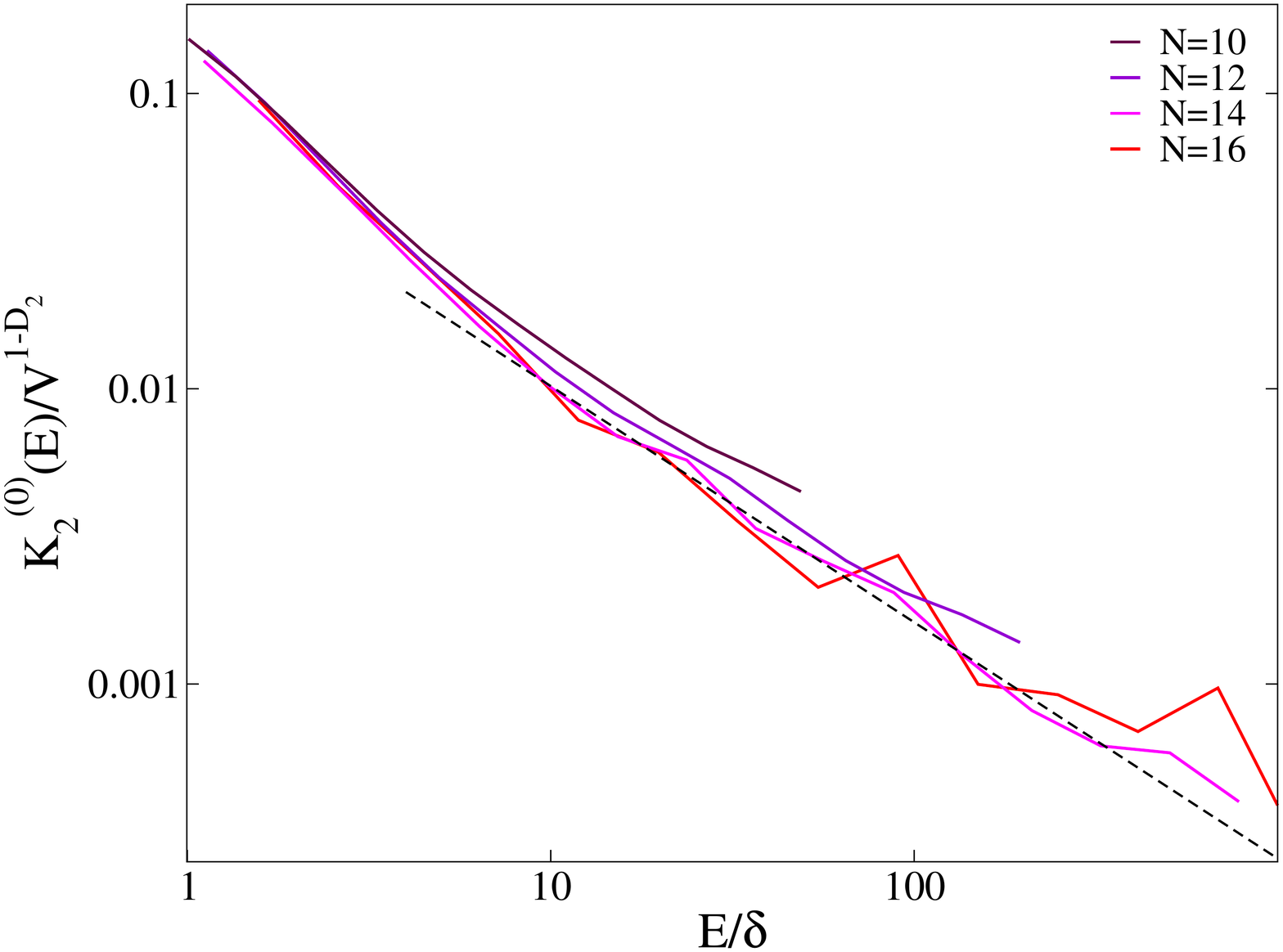}%
\caption{\label{fig:K2_scaling}
        Collapse of the data at $W=12$ obtained when rescaling the energy by the mean level spacing $\delta$ and $K_2^{(0)} (E)$ by ${\cal V}^{1-D_2}$, with $D_2 \approx 0.18$. The dashed black line is a fit of the power-law behavior observed at moderate energy separation as $K_2^{(0)} (E) \sim 1/E^{1-\kappa}$, with $\kappa \approx 0.2$, see fig.~\ref{fig:dyn}.}
\end{figure}

The change of behavior of $K_2^{(0)} (E)$ in the glassy and normal regimes of the delocalized phase is clearly visible in fig.~\ref{fig:K2}: For $W=4<W_T$ the overlap correlation function approaches a size-independent value of order $1$ when the energy is of order $\delta$.
For $W=12>W_T$, instead, the value reached by $K_2^{(0)} (E)$ at small energy separation grows with the system size.
As shown in Fig.~\ref{fig:K2_scaling}, the curves obtained for different $N$ collapse when the energy is rescaled by the mean level spacing $\delta$ and $K_2^{(0)} (E)$ by ${\cal V}^{1-D_2}$, with a spectral fractal dimension $D_2 \approx 0.18$.\cite{Bethe} Moreover, 
the overlap correlation function decays as a power-law, $K_2^{(0)} (E) \sim E^{-\gamma}$, with $\gamma \approx 1$ at small energy separation ($E \lesssim 4 \delta$) and $\gamma \approx 1 - \kappa \approx 0.8$ at moderate energy separation, consistently with the exponent which describes the decay in time of $C(t)$ (see previous section).

Interestingly, it was shown\cite{luitz_barlev} that systems that are asymptotically in a thermal state, yet exhibit anomalous relaxation and subdiffusion, must satisfy a 
modified version of the Eigenstate Thermalization Hypothesis ansatz\cite{ETH} for the off-diagonal matrix elements of local operators $\langle \alpha \vert O  \vert \beta \rangle$. Taking our definition~(\ref{eq:Oproxy}) for Bethe lattice proxies of local observables in real space, such off-diagonal elements, written in the eigenbasis of the Hamiltonian, read:
\[
\langle \alpha \vert O  \vert \beta \rangle = \sum_{x = 1}^{\cal V} \vert \langle \alpha \vert x \rangle \vert \, \vert \langle x \vert \beta \rangle \vert f(x) \, .
\]
In Ref.~[\onlinecite{luitz_barlev}] a general connection between the scaling of the variance of this object (which is tightly related to the overlap correlation function introduced above) and the non-exponential decay of dynamical correlation functions 
was derived. In particular, it was found that for subdiffusively systems the variance exhibits an anomalously slow scaling with system size than expected for diffusive systems, which corresponds to the singular behavior of $K_2^{(0)} (E)$ discussed above. Within our interpretation in terms of the freezing transition of DPRM, such unusual scaling is tracked back to the ramified structure of the wave-functions in the Fock space within the delocalized glassy phase.\cite{caveat}

\section{The effect of loops: Cayley tree vs Random Regular Graph} \label{sec:loops}
All the results presented so far have been obtained in absence of loops, i.e., by considering the loop-less
Cayley tree as the underlying lattice mimicking the Fock space. 
This has the advantage that the mapping to DPRM can be carried out without resorting to any approximation, the DP average free-energy, Eqs.~(\ref{eq:DPRMphi}) and~(\ref{eq:DPRMphiann}), is well defined and can be computed with very high numerical precision, and the 
transition between the two delocalized phases taking place at $W_T$ can be established accurately. 
However, the configuration space of $N$-body systems is more appropriately represented by RRGs, which do not posses boundaries (differently from
the Cayley tree, all site of the RRG are statistically equivalent after averaging over the disorder) and have loops at all scale (whose typical length is of
order $\ln {\cal V} \propto N$).
A crucial question that naturally arises is therefore to what extent the scenario discussed above is modified when the effect of these loops is taken into account.

\begin{figure}
\includegraphics[width=0.5\textwidth]{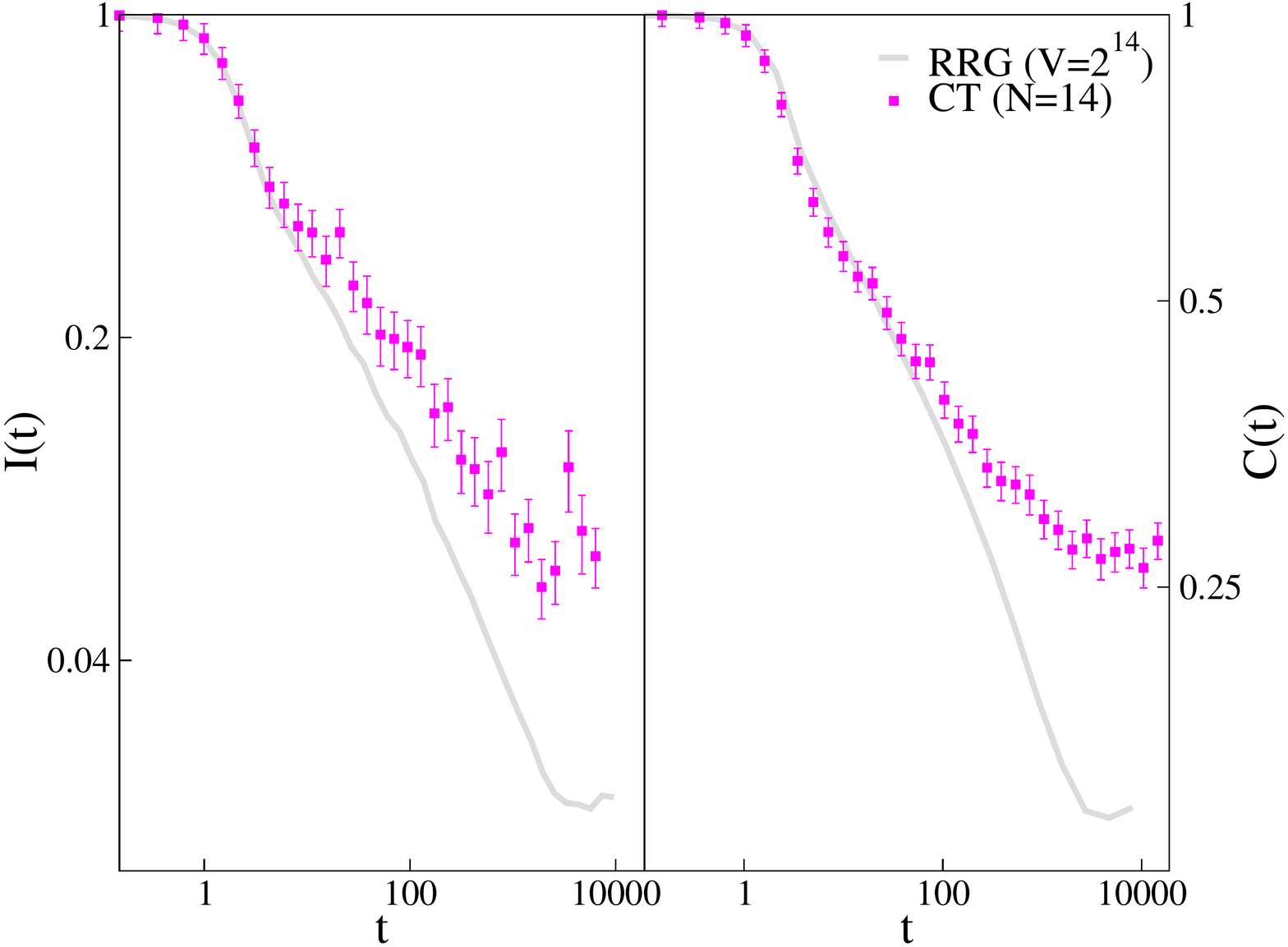}%
\caption{\label{fig:CTvsRRG}
	$I(t)$ (left panel) and $C(t)$ (right panel) as a function of $t$ for 
	$W=12$. The magenta squares show the data for Cayley trees of $N=14$ generations while the gray curve corresponds to the numerical results 
	previously reported in Ref.~[\onlinecite{PLMBL}] when using RRGs of ${\cal V} = 2^{14}$ sites as the underlying lattice.}
\end{figure}

A first piece of the answer can be obtained by comparing the quantum dynamics on the Cayley tree and on the RRG, recently analyzed in Refs.~[\onlinecite{PLMBL}] and~[\onlinecite{scardicchio_dyn}], at the same disorder strength. This comparison is shown in Fig.~\ref{fig:CTvsRRG}.
The plots indicate that 
for moderately large times and moderately large system sizes, the time dependence of the imbalance and of the correlation function on the Cayley tree is very similar to the one previously found on the RRG\cite{PLMBL} (shown in gray in  fig.~\ref{fig:CTvsRRG}).
In particular for $W=12$ we find very similar apparent exponents (within our numerical accuracy) describing the power-law decay of the imbalance and of the correlation 
function as $I(t) \sim t^{-\zeta}$ and $C(t) \sim t^{- \kappa}$, $\zeta \approx 0.4$ and $\kappa \approx 0.2$, as the ones reported in Ref.~[\onlinecite{PLMBL}] for the RRG.\cite{exponents}
However, as mentioned above, on longer time scales the apparent power-laws observed on the RRG in Ref.~[\onlinecite{PLMBL}] are actually cut-off and replaced
by an exponential decay.\cite{scardicchio_dyn,mirlintikhonov} This occurs on a timescale which
grows very fast as the localization transition is approached, as $\tau_{\rm ergo} \sim e^{B/\sqrt{W_L-W}}$,\cite{tikhonov_critical} 
and is very large already far from $W_L$. One then needs to simulate
very large samples to observe the crossover from the algebraic decay to the (stretched) exponential one.
We argue that this is due to the fact that the single-particle Anderson model on the RRG becomes eventually fully ergodic on a characteristic volume
which diverges exponentially as one approaches $W_L$ as ${\cal V}_{\rm ergo} \sim e^{A/\sqrt{W_L-W}}$.\cite{SUSY,alternative,mirlin,lemarie,Bethe,mirlintikhonov}
Instead the Anderson model on the Cayley tree displays a genuine non-ergodic behavior, with 
multifractal wave-functions in the whole delocalized phase.\cite{garel,mirlinCT,ioffe3,Bethe}

Putting all these observations together, one then comes to the following physical interpretation: On finite time scales the dynamical evolution can only explore finite regions of the
Bethe lattice. Since RRGs look locally as loop-less trees, it is natural to expect that 
the dynamics on RRGs at moderately large times is well described by the dynamics on Cayley trees.
Moreover, RRGs of moderately large sizes, i.e., smaller than the correlation volume ${\cal V}_{\rm ergo}$,
do not possess loops that are large enough to restore the ergodicity and behave {\it as if} they were in a delocalized non-ergodic phase
for all practical purposes.\cite{Bethe}
Hence the dynamical behavior on RRGs smaller than ${\cal V}_{\rm ergo}(W)$ is essentially the same as the one on the Cayley tree.
The difference between the two lattices can only be seen at very large sizes and very large times (${\cal V} > {\cal V}_{\rm ergo}$ and $t>\tau_{\rm ergo}$): On the Cayley tree the power-law regimes will persist
up to arbitrary large times, while on the RRG they are cut-off on a huge crossover scale. In other words, while a sharp transition
takes place at $W_T$ in the limit of infinite Cayley trees, this transition becomes only a crossover on RRGs, and is smeared 
out if ${\cal V} \gg {\cal V}_{\rm ergo}$.

\begin{figure}
\includegraphics[width=0.46\textwidth]{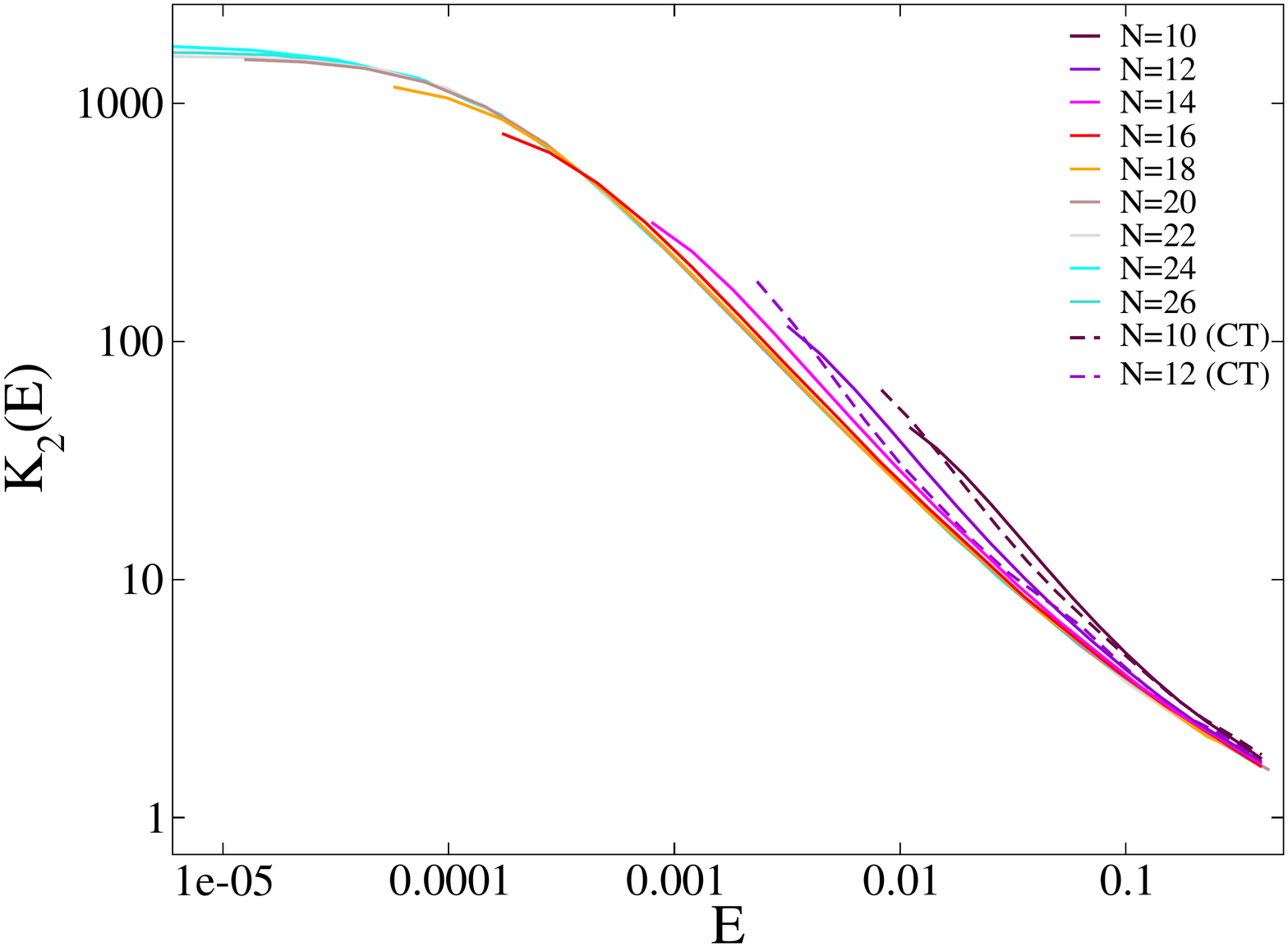}%
\caption{\label{fig:K2RRG}
        Overlap correlation function $K_2(E)$ as a function of $E$ for RRGs of several system sizes (${\cal V} = 2^N$, with $N=10, \ldots, 26$) obtained with the approximate technique described in Ref.~[\onlinecite{Bethe}] for $W=12$. 
        The overlap correlation function on RRGs of size smaller than the correlation volume ${\cal V}_{\rm ergo}$ is very similar to the one found on the root of Cayley trees of (about) the same size (dashed curves).}
\end{figure}

In order to understand the effect of loops on the quantum dynamics, one can also contrast the properties of the overlap correlation function $K_2^{(0)} (E)$ on the Cayley trees discussed in the previous section with the ones found on RRGs of (about) the same sizes and at the same disorder strength.\cite{Bethe,mirlintikhonov} In fig.~\ref{fig:K2RRG} we show the overlap correlation function $K_2(E)$ computed on the RRG\cite{K2RRG} with the approximate technique of Ref.~[\onlinecite{Bethe}] for $W=12$ 
and for several system sizes ${\cal V} = 2^N$, with $N=10, \ldots, 26$. For 
system sizes smaller than the ergodic crossover scale ${\cal V}_{\rm ergo}$ (e.g., ${\cal V} \lesssim 2^{16}$) the overlap correlation function on the RRG behaves very similarly to the Cayley tree. However, for larger system sizes the dependence of $K_2(E)$ on the volume saturates and the curves converge to a size-independent limiting non-singular function
characterized by a plateau at small energy followed by a fast decrease at larger energy. As discussed above, this is the typical metallic behavior found on the (fully ergodic) delocalized side of the Anderson transition.
The energy scale $E_{\rm Th}$ over which the plateau extends stays finite in the thermodynamic limit and
represents the width of the energy band within which GOE-like correlations are established.\cite{thouless}

These findings confirm the scenario discussed in the previous section: RRGs smaller than the ergodic crossover scale behaves for all practical purposes as (the bulk of) Cayley trees, while full ergodicity and standard metallic behavior is recovered for larger samples on an energy scale $E_{\rm Th}$ which vanishes exponentially fast close to the Anderson localization.\cite{Bethe,mirlintikhonov} From Eq.~(\ref{eq:CtFT}) one then expects a crossover to a standard exponential decay of dynamical correlations on RRGs of size larger than ${\cal V}_{\rm ergo}$ and on time scales larger than $\tau_{\rm ergo} = \hbar/E_{\rm Th}$.\cite{mirlintikhonov}
Conversely on Cayley trees $K_2^{(0)} (E)$ is characterized by a genuinely singular limit in the whole delocalized glassy phase, and the power-laws in the dynamics persist to infinitely long times.

A more detailed discussion of the crossover phenomena associated to the non-ergodic-like regime in RRGs can be found in Ref.~[\onlinecite{Bethe}].

\section{Conclusions and perspectives} \label{sec:conclusions}

In conclusion, using the non-interacting Anderson model on the Cayley tree 
as a toy model for the quantum
many-body dynamics,\cite{dot,BetheProxy1,BetheProxy2,scardicchioMB,PLMBL} we have proposed a transparent theoretical explanation of the subdiffusive behavior and anomalously slow relaxation observed on the thermal side of isolated disordered many-body systems, in terms of delocalization along rare and ramified paths in the Fock space.
In particular, we have 
shown the existence of a glass transition (in the thermodynamic limit) 
separating two different extended phase: A metallic-like phase at weak disorder ($0 \le W \le W_T$) where delocalization occurs on an exponential number of paths, 
and a bad metal-like phase at intermediate disorder ($W_T \le W \le W_L$) where delocalization takes place only on few, specific, 
ramified paths.
Translating our results to the many-body problem, this means that at weak disorder the number of site orbitals in the Fock space
to which the initial state is effectively coupled grows exponentially with the distance,\cite{exponential} while 
in the intermediate glassy phase resonances are formed only on rare
site orbitals 
on very distant generations, implying that energy and spin transport is highly heterogeneous, precisely as predicted in the pioneering work of Ref.~[\onlinecite{dot}].

The physical interpretation of the unusual relaxation observed in the bad metal phase emerging from this picture is complementary to the Griffiths one.
In both scenarios subdiffusion and non-exponential relaxation are the result of the presence of heavy-tailed distribution
(which are often associated to the failure of the central limit theorem).
However, according to the Griffiths picture, the unusual slow dynamics follows from exponentially rare inclusions {\it in real space} which acts as
kinetic bottlenecks and yield effective barriers with exponentially large relaxation times.\cite{griffiths,griffiths2,deroeck} 
According to the perspective proposed here, instead, the slow dynamics is tracked back to very heterogeneous delocalization of local excitations along rare, 
disorder-dependent paths {\it in Fock space} with a singular and heavy-tailed probability distribution of the decorrelation probability.
In many practical situations (especially for systems of limited size) both effects could be at play simultaneously.
Studying the behavior of typical versus average correlation functions, as recently done in Ref.~[\onlinecite{typvsav}] for a Floquet model, should be a good probe to distinguish and disentangle them.

Notice that within our approach the unusual slow dynamics and power-law behavior 
is {\it not} just related to the non-ergodicity of wave-functions. In fact, the eigenfunctions of the non-interacting Anderson model on the 
Cayley tree are multifractal in the whole delocalized phase,\cite{garel,mirlinCT,exponential,Bethe} whereas the dynamical observables display a fast exponential 
decay for $W < W_T$, as shown in fig.~\ref{fig:dyn}. The emergence of algebraic decays is instead associated to the freezing glass transition of the 
delocalizing paths and to the singular statistical properties of the decorrelation probability along these paths.
This can be understood in terms of the depinning transition of the DP in the glassy phase,\cite{gabriel} which yield abrupt rearrangements of the preferred conformations of the delocalizing paths when a parameter like the energy is varied. This results in a singular behavior of the overlap correlation function between eigenstates at different energies, which is essentially the Fourier transform to frequency domain of the dynamical correlation function.

Of course, our toy model is the result of a series of extreme (over)simplifications. More work is needed to go beyond the approximations considered here and establish a precise
connection between the spreading of the wave-packet on a complicated graph and the many-body dynamics of a realistic system.
In this respect it would be important to establish a direct quantitative relationship between the exponents describing the power-law decay
of the dynamical observables such as the Bethe lattice proxies of the imbalance and of the equilibrium correlation function\cite{PLMBL} and
the statistical properties of the decorrelation probability along the paths, which becomes singular above $W_T$.
It would be also interesting to understand how the properties of the intermediate glassy phase depend on the connectivity of the lattice (which, to mimic 
the structure of the configuration space of a $N$-body interacting system should increase as $N$), and on the scaling of the random on-site energies
(which should be thought as the counterpart of {\it extensive} energies of the $N$-body system). It is in fact well established that the localization
transition scales as $W_L \sim k \ln k$,\cite{ATA,victor,ioffe3} and it would be interesting to check whether $W_T$ follows the same asymptotic behavior 
or a different one. (A naive estimation of $W_T$ obtained using the forward-scattering approximation on the Bethe lattice seem to suggest that $W_T \propto \sqrt{k}$.)
A step forward in these directions would be to adapt the present analysis to the quantum version of the REM, which is possibly the simplest system exhibiting a MBL transition,\cite{pal}
as it allows to retain the local connectivity of the configuration space and the scaling with $N$ of the random energies, and yet to neglect the
correlations between random energies on different site orbitals.
Finally, the glass transition of the DP analyzed here
takes place as a front propagating from the root of the tree towards the boundary, as already noticed in Ref.~[\onlinecite{mirlinCT}]. It would be
interesting to study the extended phase diagram of the problem as a function of the dimensionless distance from the root which, in terms of
dynamical evolution, is akin to time.

A crucial point, partially discussed here, concerns to what extent the details of the specific structure of the underlying lattice mimicking the configuration space are important. In particular in this paper we have focused on the effects of loops, which are completely disregarded 
when one considers loop-less
Cayley trees as the underlying lattice for our toy model. Doing this has the advantages
that the mapping to DPRM can be carried out without resorting to any approximation and that the glass transition found at $W_T$ survives in the thermodynamic limit. 
However, the configuration space of $N$-body systems is more appropriately represented by RRGs, which do not posses boundaries (differently from
the Cayley tree, all site of the RRG are statistically equivalent after averaging over the disorder) and have loops at all scale (whose typical length is of
order $\ln {\cal V} \propto N$).
However, as discussed above, full ergodicity is (most likely) eventually recovered in the whole delocalized phase of the Anderson model on the RRG for system larger than a correlation volume which diverges exponentially (as ${\cal V}_{\rm ergo} \sim e^{A/\sqrt{W_L-W}}$, see Refs.~[\onlinecite{SUSY}], [\onlinecite{tikhonov_critical}], and~[\onlinecite{alternative}]) approaching the Anderson transition.\cite{mirlin,lemarie,levy,Bethe,mirlintikhonov} 
One then expects that the apparent power-law decay of dynamical correlation functions observed on the RRG should be eventually cut-off for sizes larger 
than ${\cal V}_{\rm ergo}$, 
and should be replaced by a standard exponential decay.\cite{scardicchio_dyn,mirlintikhonov} Yet, the time scale at which the decay of the correlation functions can be distinguished by an algebraic one
also diverges exponentially (as $\tau_{\rm ergo} = \hbar/E_{\rm Th} \sim e^{B/\sqrt{W_L-W}}$) approaching the Anderson transition, and is already very large far from it. 
Since on large but finite times the dynamics can only explore a large but finite volume, close enough to the
localization transition, the dynamics of the system
is slow and unusual for many decades (and well described by the Cayley tree), although it becomes eventually ergodic at large times.

The Fock space of a realistic many-body Hamiltonian, such as the disordered spin chain of Eq.~(\ref{eq:HMB}), is a $N$-dimensional hyper-cube, with the extra complication that on-site random energies are strongly correlated. An interesting possibility would be then to diagonalize numerically the many-body Hamiltonian and analyze the statistical properties of the delocalizing pahts in the configuration space (similarly to the recent analysis of~[\onlinecite{laflorencie}]) together with the scaling behavior of the spectral probe $K_2(E)$ as a function of the system size, and benchmark the results onto the quantum (equilibrium and out-of-equilibrium) many-body dynamics. This opens a new theoretical perspective to investigate the MBL transition and to characterize the properties of the bad metal phase.

\begin{acknowledgments}
We acknowledge support from the Simons Foundation (\#454935, Giulio Biroli).
\end{acknowledgments}

\appendix

\section{Spectral representation of ${\cal W} (x^{(N)})$} \label{app:spectral}

In order to obtain the spectral representation of ${\cal W} (x^{(N)})$ in terms of the elements of the resolvent matrix one needs to define the following correlation function:
\[
\begin{aligned}
\gamma(x,E) = \left \langle \sum_\alpha \left \vert \langle x \vert \alpha \rangle \right \vert^2 \left \vert \langle \alpha \vert 0 \rangle \right \vert^2 \delta (E - E_\alpha) \right \rangle \, ,
\end{aligned}
\]
and the local DoS on a site $x$: 
\[
\rho_x (E) = \sum_\alpha \left \vert \langle \alpha \vert x \rangle \right \vert^2 \delta (E - E_\alpha) \, ,
\]
whose spectral representations are simply given by:
\[
\begin{aligned}
\gamma(x,E) & = \lim_{\eta \to 0^+} \frac{\eta \vert {\cal G}_{0,x} (E) \vert^2}{\pi \rho (E)} \, , \\
\rho_x (E) & = \lim_{\eta \to 0^+} \frac{{\rm Im} {\cal G}_{x,x} (E)}{\pi} \, ,
\end{aligned}
\]
where $\rho(E) = (1/{\cal V}) \sum_x \rho_x (E)$ is the total DoS.
The infinite time limit of the wave-function amplitude on the boundary site $x^{(N)}$ starting from the root of the tree can be then written as:
\[
\begin{aligned}
{\cal W} (x^{(N)}) & 
\equiv \lim_{t \to \infty} \left \vert \langle x^{(N)} \vert \psi (t) \rangle \right \vert^2 
= \frac{\sum_{\alpha}^\star \vert \langle x^{(N)} \vert \alpha \rangle \vert^2  \left \vert \langle \alpha \vert 0 \rangle \right \vert^2}
{\sum_\alpha^\star \vert \langle \alpha \vert 0 \rangle \vert^2} \\ 
& \qquad \qquad = \frac{\int_{-\Delta E}^{+ \Delta E} \rho (E) \gamma(x,E) {\rm d} E}
{\int_{-\Delta E}^{+ \Delta E} \rho_0 (E) {\rm d} E} \, .
\end{aligned}
\]
Assuming that $\Delta E$ is small enough such that the dependence of $\gamma(x,E)$ and $\rho_0(E)$ on the energy is weak, one can approximate the integrals by the values at the middle of the band, which finally yields the approximate expression given in Eq.~(\ref{eq:amplitude}).

\section{Relationship between ${\rm Im} {\cal G}_{0,0}$ and the partition functions of directed polymers on the Cayley tree} \label{app:ZvsImG}
The recursive equation~(\ref{eq:recursion}) for the imaginary part of the cavity Green's functions can be telescoped as:
\begin{equation} \label{eq:DPRMImG}
        {\rm Im} G_{x \to y} = | G_{x \to y}|^2 \left( \eta +
        \!\!\! \sum_{z \in \partial x / y}  \!\!\!\! {\rm Im} G_{z \to x} \right) \, .
\end{equation}
From this equation, the Green's function at the root of the tree can be re-expressed as:
\begin{equation} \label{eq:DPRMpaths}
	\begin{aligned}
		&{\rm Im} {\cal G}_{0,0} = \eta \left | {\cal G}_{0,0} \right |^2 \sum_{M=0}^{N-2} 
		\sum_{{\rm{\sf P_M}}} \prod_{i=1}^{M} \left | G_{x^{(i)}
        \to x^{(i-1)}} \right |^2 \\ 
		& \qquad + \left | {\cal G}_{0,0} \right |^2 \sum_{{\rm{\sf P_{N-1}}}} \left ( \prod_{i=1}^{N-1} \left | G_{x^{(i)}
        \to x^{(i-1)}} \right |^2 \right) {\rm Im} G_{x^{(N)} \to x^{(N - 1)}}	
        \end{aligned}
\end{equation}
where the sums are over all directed paths ${\rm {\sf P_M}}$ of length $M$ connecting the root of the tree with the sites of
the $M$-th generation, 
and $x^{(i)} \to x^{(i - 1)}$ are all the edges belonging to ${\rm {\sf P_M}}$ connecting
the site $x^{(i)}$ of the $i$-th generation to the site $x^{(i-1)}$ of the $(i-1)$-th generation.
The first part of the r.h.s. of Eq.~(\ref{eq:DPRMpaths}) can be interpreted as the sum of ($\eta$ times) the partition functions of DPs of length $M$
originating from the root of the Cayley tree in presence of the quenched random energy landscape generated by the $| G_{x^{(i)} \to x^{(i-1)}}|^2$.
Since the cavity Green's function on site $x^{(N)}$ of the boundary of the tree in absence of its only neighbor $x^{(N-1)}$ is simply
$G_{x^{(N)} \to x^{(N - 1)}} = (\epsilon_{x^{(N)}} - i \eta)^{-1}$, one has that ${\rm Im} G_{x^{(N)} \to x^{(N - 1)}} = \eta |G_{x^{(N)} \to x^{(N - 1)}}|^2$.
Thus the second line of the r.h.s. of Eq.~(\ref{eq:DPRMpaths}) exactly coincides with the r.h.s. of Eq.~(\ref{eq:Pdecorr}).
One then obtains Eq.~(\ref{eq:ImGZpre}) of the main text.

The term $M=N$ of~(\ref{eq:DPRMpaths}) 
gives the leading contribution to ${\rm Im} {\cal G}_{0,0}$, as also
confirmed by fig.~\ref{fig:distrib}, which shows that asymptotically $P(\eta Z_{\rm DP} (N)) \sim Q ({\rm Im} {\cal G}_{0,0})$.

This is the result that one would obtain neglecting the imaginary regulator inside the barkets of Eq.~(\ref{eq:DPRMImG}) from the beginning.
In fact, the natural scale of the imaginary regulator is the mean level spacing, $\eta = c \delta$, with $\delta = 1/({\cal V} \rho_0)$. 
Hence $\eta$ behaves as $W/{\cal V}$, while the typical value of ${\rm Im} G_{x^{(i)} \to x^{(i-1)}}$ grows under iteration 
in the delocalized phase.~\cite{ImGgrows}
This result has an obvious physical interpretation: 
the spreading of the level width of local exciatations (described essentially by ${\rm Im} {\cal G}_{0,0}$) is tightly related to the 
probability that such excitations travel far away on the tree.


\section{Relationship between $\phi(\beta = 1)$ and the Lyapunov exponent} \label{app:lyap}
Let us focus on the iteration relations describing the propagation of the imaginary part of the cavity Green's function from the leaves to the root of a Cayley tree of $N$ generations. 
As mentioned above, on the boundary of the tree ${\rm Im} G_{x^{(N)} \to x^{(N - 1)}} =  \eta/(\epsilon_{x^{(N)}}^2 + \eta^2)$.
The typical value of ${\rm Im}G$ is thus of order $\eta$, 
${\rm Im}G^{\rm typ}_N = e^{\langle \ln {\rm Im} G_{x^{(N)} \to x^{(N - 1)}} \rangle} \approx \eta$. 
In the whole delocalized phase ${\rm Im}G^{\rm typ}$ grows (by definition) under iteration.
Such growth can be characterized by a Lyapunov exponent $\Lambda (W)$ as 
${\rm Im}G^{\rm typ}_{i-1} = e^{\Lambda} {\rm Im}G^{\rm typ}_i$.
$\Lambda(W)$ is positive in the delocalized phase, decreases as $W$ is increased, 
and vanishes at the Anderson localization transition at $W_L$.\cite{ioffe1,ioffe3,ATA,aizenmann} 
In the large $N$ limit one then has that:
\begin{equation} \label{eq:lyap}
	{\rm Im} {\cal G}^{\rm typ}_{0,0} \approx e^{N \Lambda} {\rm Im} G^{\rm typ}_N \approx \eta e^{N \Lambda} \, .
	\end{equation}
Moreover, since ${\rm Im} {\cal G}_{0,0} \simeq \eta Z_{\rm DP}$ (see App.~\ref{app:ZvsImG} and fig.~\ref{fig:distrib}), we have:
\[
	\phi(\beta) \simeq \frac{1}{\beta N} \left \langle \ln \left ({\rm Im} {\cal G}_{0,0}/\eta \right)^\beta
	\right \rangle \, ,
\]
which implies that:
\begin{equation} \label{eq:phi_lyap}
{\rm Im} {\cal G}^{\rm typ}_{0,0} \simeq \eta e^{N \phi (\beta = 1)} \, .
\end{equation}
In conclusion, from Eqs.~(\ref{eq:lyap}) and (\ref{eq:phi_lyap})  one obtains that in the thermodynamic limit $\phi (\beta = 1)$ coincides asymptotically with
the Lyapunov exponent $\Lambda (W)$. For $W>W_T$ we find that $\beta_\star < 1$. Since $\phi(\beta)$ remains constant for $\beta > \beta_\star$, one has
that in the intermediate phase $\phi(\beta = 1) = \phi(\beta_\star)$. We indeed find that at the localization transition $\phi (\beta_\star) = 0$
and $\beta_\star = 1/2$ 
as rigorously proved in Refs.~[\onlinecite{aizenmann}] and~[\onlinecite{aiz_war}] (and indirectly found in Ref.~[\onlinecite{ATA}], see also Refs.~[\onlinecite{garel}], [\onlinecite{noi}], and~[\onlinecite{levy}] for further details).

\section{Return probability} \label{app:return}
The correlation function defined in Eq.~(\ref{eq:Ct}) is actually equivalent to
the so-called return probability, recently studied in Ref.~[\onlinecite{scardicchio_dyn}] on the RRG.
The return probability (constructed using eigenstate within the bandwidth $[-\Delta E, \Delta E]$) is defined as:
\[
	\begin{aligned}
		P_R (t) & = 
	\left \vert \langle 0 \vert \psi (t) \rangle \right \vert^2 = \left \vert \langle 0 \vert e^{- i {\cal H} t / \hbar} \vert 0 \rangle \right \vert^2 \\
		& = \frac{\sum_{\alpha,\beta}^\star \left \vert \langle \alpha \vert 0 \rangle \right \vert^2 \left \vert \langle \beta \vert 0 \rangle 
		\right \vert^2 e^{-i (E_\alpha - E_\beta) t / \hbar}}
		{\left (\sum_{\alpha}^\star \left \vert \langle 0 \vert \alpha \rangle \right \vert^2 \right )^2}\, .
	\end{aligned}
\]
(The normalization factor in the denominator ensures that $P_R(0)=1$.)

\begin{figure}
\includegraphics[width=0.48\textwidth]{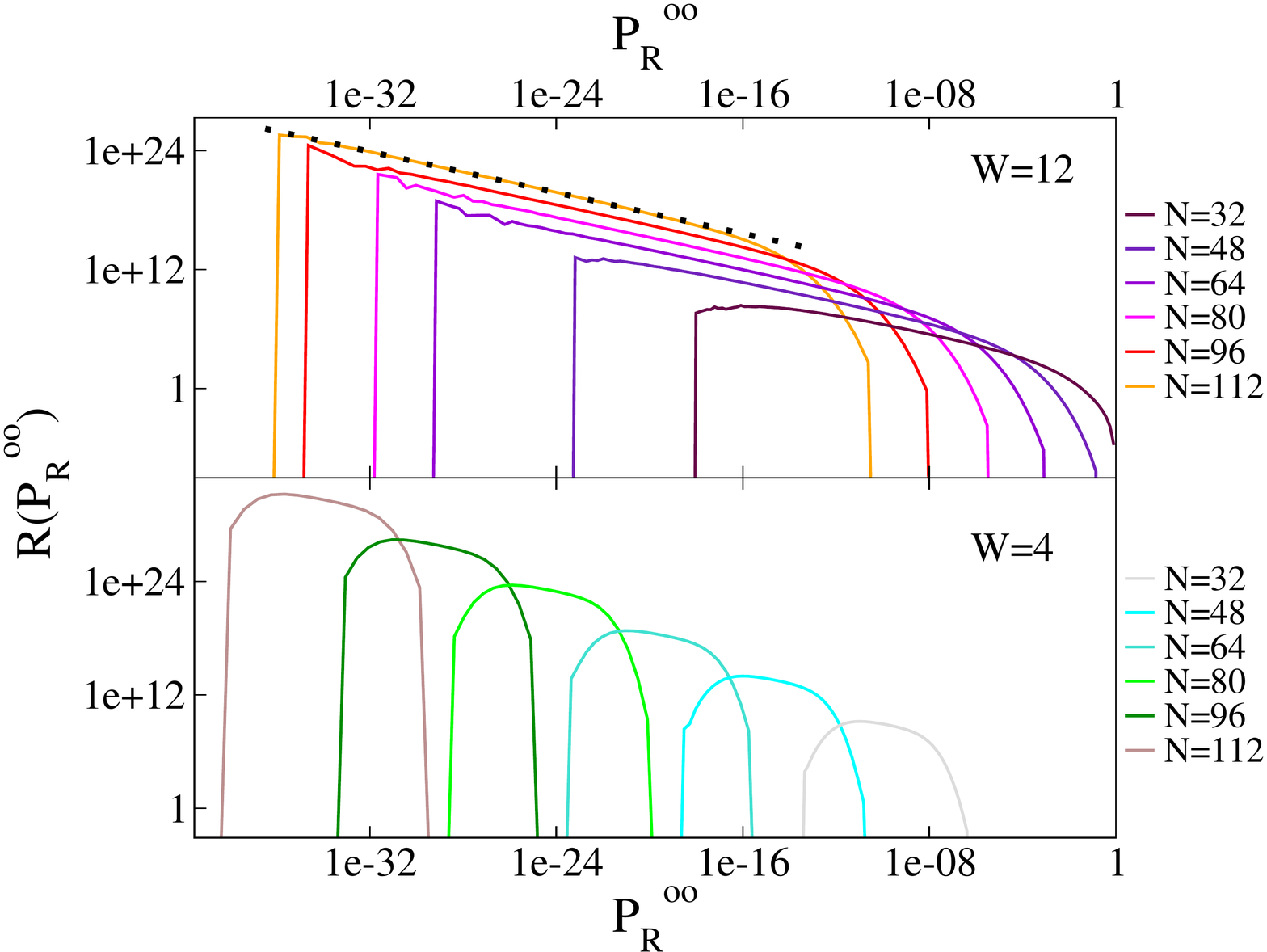}%
\caption{\label{fig:PR}
        Log-log plot of the probability distributions of the long-time limit of the return probability $R(P_R^\infty)$ 
	for Cayley trees of $N$ generations (with $N$ going from $32$ to $112$), and for $W = 12 > W_T$ (top panel) and $W = 4 < W_T$ (bottom panel).
	The black dotted straight line in the top panel corresponds to a power-law fit of the data as $R(P_R^\infty) \sim (P_R^\infty)^{- \tilde{D}_2}$
	with $\tilde{D}_2 \approx 0.52$.}
\end{figure}

In the long time limit $P_R^\infty \equiv P_R(t \to \infty)$ measure the probability that the system stays localized around the root of the tree $|0 \rangle$ and keeps
memory for infinite time of the initial configuration,
\[
	P_R^\infty = \frac{\sum_\alpha^\star \left \vert \langle \alpha \vert 0 \rangle \right \vert^4}
	{\left (\sum_{\alpha}^\star \left \vert \langle 0 \vert \alpha \rangle \right \vert^2 \right )^2}\, ,
\]
and decays to zero with ${\cal V}$ in the whole delocalized phase.
Using the expressions of the correlation function $\gamma(x,E)$ and of the LDoS $\rho_x (E)$ introduced in App.~\ref{app:spectral}, $P_R^\infty$ can be rewritten as:
\[
P_R^\infty = \frac{\int_{-\Delta E}^{+ \Delta E} \rho(E) \gamma(0,E) {\rm d} E}{\left ( \int_{-\Delta E}^{+ \Delta E} \rho_0 (E) {\rm d} E \right )^2} \, .
\]
Assuming once again that $\Delta E$ is small enough such that the dependence of $\gamma(0,E)$ and $\rho_0(E)$ on the energy is weak (and assuming that $\rho_0 (0) \simeq \rho (0)$), and using
the spectral representation of $\gamma(0,E)$ one finally obtains:
\[
	P_R^\infty \approx \lim_{\eta \to 0^+} \frac{\eta \vert {\cal G}_{0,0} \vert^2}{2 \Delta E \rho_0^2} 
	\, .
\]
The probability distributions of the long time limit of the return probability, $R(P_R^\infty)$, are plotted in fig.~\ref{fig:PR} for several system sizes ($N=32,\ldots,112$) for
two values of the disorder strength ($W=4$ and $W=12$), showing that they behave in a completely different way on the two sides of the transition:
For $W=4<W_T$ the probability distributions of the long time limit of the return probability follows a trivial 
scaling behavior, $R_N(P_R^\infty) = {\cal V} R_\infty ({\cal V} P_R^\infty)$, where $R_\infty (x)$ is a narrow function which decays fast to zero.
Hence both the typical value and the average value of $P_R^\infty$ go to zero with the same exponent as ${\cal V}^{-1}$.
Conversely, for $W=12>W_T$ the probability distributions $R_N(P_R^\infty)$ are multifractal (i.e., do not follow a simple scaling behavior) and are 
characterized by heavy-tails 
which correspond to anomalously large rare values of $P_R^\infty$.
The average and the typical value go to zero as $\langle P_R^\infty \rangle \sim {\cal V}^{- \tilde{D}_2}$ and $e^{\ln \langle P_R^\infty \rangle} \sim {\cal V}^{- \tilde{D}_2^{\rm typ}}$
with the exponents $\tilde{D}_2 \approx 0.514$ and $\tilde{D}_2^{\rm typ} \approx 0.517$ both smaller than one.  
(The fractal exponent $D_1$ describing the scaling of the typical value of the imaginary part of the Green's function at the root of the Cayley tree as 
$e^{\langle \ln {\rm Im} {\cal G}_{0,0} \rangle } \sim {\cal V}^{- D_1}$ is $D_1 \approx 0.483$ for $W=12$.)
This is a manifestation of the non-ergodicity of the wave-functions.\cite{mirlinCT} 
A similar result has also been recently found for the Anderson model on the RRG.\cite{scardicchio_dyn}
However, repeating the same analysis on the RRG we find that in this case the non-ergodic behavior is established only for systems smaller than the 
correlation volume, 
while full ergodicity is restored in the limit of very large samples (i.e., $\tilde{D}_2=\tilde{D}_2^{\rm typ} = 1$ in the whole delocalized phase provided that
${\cal V} \gg {\cal V}_{\rm ergo}$). See Refs.~[\onlinecite{Bethe}] and~[\onlinecite{mirlintikhonov}] for more details.

In a similar way, one can define the probability that a particle that sits at the root of the tree at time $0$ is found on the boundary of the tree
at time $t$ (which is the situation analyzed in the main text).
Using the fact that the wave-functions' amplitudes can be written in terms of the matrix elements of the resolvent as 
$\vert \langle x \vert \psi (t \to \infty ) \rangle \vert^2 \approx \lim_{\eta \to 0^+} \eta \vert {\cal G}_{x,0} \vert^2/{\rm Im} {\cal G}_{0,0}$, 
using Eq.~(\ref{eq:corr}) to express the correlation functions ${\cal G}_{x,0}$ in terms of the cavity Green's functions on the edges
connecting sites $0$ and $x$, and recalling Eq.~(\ref{eq:ImGZpre}), one finally obtains:\cite{PRinfvsIPR}
\[
        \begin{aligned}
                & P_B (t) =
		\sum_{x^{(N)}} \left \vert \langle x^{(N)} \vert \psi (t) \rangle \right \vert^2 = \sum_{x^{(N)}} \left \vert \langle x^{(N)} \vert e^{- i {\cal H} t / \hbar} \vert 0 \rangle \right \vert^2 \\
		& \,\,\,\,\,\,\,\, = \sum_{x^{(N)}} \sum_{\alpha,\beta}^\star \langle \alpha \vert 0 \rangle \langle x^{(N)} \vert \alpha \rangle 
		\langle \beta \vert 0 \rangle \langle x^{(N)} \vert \beta \rangle \, 
                e^{-i (E_\alpha - E_\beta) t / \hbar} \, .
        \end{aligned}
\]
The long time limit of this object, defined in Eq.~(\ref{eq:Pdecorr}), then reads:
\[
	P_B^\infty \approx \frac{\eta \sum_{x^{(N)}} \vert {\cal G}_{x^{(N)},0} \vert^2}{{\rm Im} {\cal G}_{0,0}} = \frac{Z_{\rm DP} (\beta=1,N)}{\sum_{M=0}^N 
	Z_{\rm DP} (\beta=1,M)}\, .
\]
The study of the numerator and the denominator {\it separately} has been the main focus of the paper (see in particular Sec.~\ref{sec:dprm}). 
As shown in the main text, both the numerator and the denominator show a freezing transition of the paths contributing to the partition functions above a critical value
of the disorder $W_T$, related to the glass transition of DPRM. The probability distributions of the numerator and of the denominator are both plotted
in Fig.~\ref{fig:distrib}, showing that they become singular and heavy-tailed for $W>W_T$.
However $P_B^\infty$ itself is insensitive to such transition: The probability distribution of $P_B^\infty$ converges to a stable non-singular limit at
large $N$ on both sides of the transition, and its average and typical values are both of $O(1)$ in the whole delocalized phase, and 
vanish at $W_L$.


\begin{thebibliography}{99}
\bibitem{NCT} In fact the total number of sites of a Cayley tree of $N$ generations and connectivity $k+1=3$ is ${\cal V} = 3 \cdot 2^{N-1} - 2$.
\bibitem{projector} We focus on a small interval $[-\Delta E,\Delta E]$ where $\Delta E$ is much smaller than the bandwidth ($\sim W$)
	and much larger than the mean level spacing, $\delta = 1/({\cal V} \rho (E=0))$. In practice we have set
	$\Delta E = 0.25$, and checked that our results do not change if the interval is increased or reduced by a factor $2$.
\bibitem{cvalue} We set $c=4$ henceforth, and we have explicitly checked that the results discussed below 
do not depend on the choice of value of the constant $c$, provided that it stays of order $1$ and does not scale with the volume ${\cal V}$.
\bibitem{exponential} Note that even in the regime $0 < W < W_T$ the paths contributing to
        the sum~(\ref{eq:Pdecorr}) are characterized by a branching number $\overline{k}$ strictly smaller than $k$,
        as already predicted in Ref.~[\onlinecite{dot}], and hence occupy a vanishing fraction of the tree as they
        are exponentially less numerous than the total number of paths.
\bibitem{exponents} The exponent describing the power-law decay of the imbalance is found to be the double of the one of the
        correlation function, consistently on what found in many-body systems,\cite{reviewMBL3} and also predicted on the basis of scaling arguments.\cite{BarLev}
\bibitem{alternative} Recently, it has been proposed~\cite{ioffe3} a different expression for ${\cal V}_{\rm ergo} \sim e^{A/(W_L-W)}$.
\bibitem{caveat} Note that, however, we do not find the same relation between the dynamical exponent $\kappa$ describing the power-law decay of the correlation function $C(t)$ and the scaling as ${\cal V}^{-D_2}$ of the variance of the off-diagonal elements $\langle \alpha \vert O \vert \beta \rangle$ as in Ref.~[\onlinecite{luitz_barlev}]. This is due to the fact that within our toy model the Thouless time scales as ${\cal V}^{1-D_2}$, while in an interacting $N$-body spin-chain  it is expected to scale as $N^{1/\kappa}$.
\bibitem{K2RRG} Since the RRG is statistically invariant by translation, the overlap correlation function can be obtained by averaging over all sites of the lattice, hence the superscript $(0)$ appearing in the definition~(\ref{eq:K2}) can be dropped.
\bibitem{ImGgrows} In particular, the typical value of the 
	imaginary part of the cavity Green's functions is of order $\eta$ on the leaves of the tree, ${\rm Im} {\cal G}_{x_N \to x_{N-1}} = \eta/(\epsilon_{x_N}^2 + \eta^2)$,
        and grows under iteration as one moves towards the center roughly as $\eta e^\Lambda$, with an exponent $\Lambda(W)$ 
	which is akin to a Lyapunov exponent: $\Lambda$ is by definition greater than zero in the delocalized phase and is
        a decreasing function of $W$, vanishing at the Anderson localization transition at $W_L$.\cite{ioffe1,ioffe3,ATA,aizenmann}
        As a result, after few iterations the typical value of ${\rm Im}G$ becomes much larger than $\eta$ and in the large $N$ limit 
	the imaginary regulator in the square brakets of Eq.~(\ref{eq:DPRMImG}) can be essentially dropped.
\bibitem{PRinfvsIPR} Note that the definition of $P_R^{\infty}$ in terms of the Green's functions is very similar but slightly different than the definition of the Inverse Participation Ratio (IPR) associated to the wave-functions' amplitude at the root of the tree.\cite{Bethe} In fact, the IPR of eigenstates around zero energy on site $0$ is $\Upsilon_2^{(0)} = \sum_\alpha \langle \alpha \vert 0 \rangle \vert^4 \delta(E_\alpha) \simeq \eta \vert {\cal G}_{0,0} \vert^2 / (\pi \rho_0)$, where $\rho_0 = (1/N) \sum_{x=1}^{\cal V} {\rm Im} {\cal G}_{0,0}$. On the RRG $P_R^{\infty}$ and  $\Upsilon_2^{(0)}$ coincide on average, since all sites are statistically equivalent and $\rho_0 = \langle {\rm Im} {\cal G}_{0,0} \rangle /\pi$, while on the Cayley tree they are in general different. Hence we will use the fractal exponent $\tilde{D}_2$ to describe the scaling of $P_R^{\infty}$ with the system size (instead of the standard $D_2$ used for the IPR).
\end{thebibliography}
\end{document}